\documentstyle[11pt]{article}
\batchmode
\begin{document}
\small
\renewcommand{\baselinestretch}{1.2}
\normalsize
\def \bc {\begin{center}}
\def \ec {\end{center}}
\def \be {\begin{equation}}
\def \ee {\end{equation}}
\def \mic { $ \mu$m }
\def \lya {Ly$\alpha$}
\def \lyb {Ly$\beta$}
\def \lyg {Ly$\gamma$}
\def \ha {H$\alpha$}
\def \hb {H$\beta$}
\def \etal {{\it et al.}}
\def\micron{\mu{\rm m}}
\def\Angstrom{\,{\rm ~\AA}}		
\def\eV{{\rm eV}}
\def\g{{\rm g}}
\def\cmM{{\rm cm}^{-1}}
\def\cmMMM{{\rm cm}^{-3}}
\def\etal{{\it et al.}}
\newbox\grsign \setbox\grsign=\hbox{$>$} \newdimen\grdimen \grdimen=\ht\grsign
\newbox\simlessbox \newbox\simgreatbox
\setbox\simgreatbox=\hbox{\raise.5ex\hbox{$>$}\llap
     {\lower.5ex\hbox{$\sim$}}}\ht1=\grdimen\dp1=0pt
\setbox\simlessbox=\hbox{\raise.5ex\hbox{$<$}\llap
     {\lower.5ex\hbox{$\sim$}}}\ht2=\grdimen\dp2=0pt
\def\gtsimeq{\mathrel{\copy\simgreatbox}}	
\def\ltsimeq{\mathrel{\copy\simlessbox}}	
\bc \Large {\bf THE ULTRAVIOLET EMISSION PROPERTIES OF
FIVE LOW-REDSHIFT ACTIVE GALACTIC NUCLEI AT HIGH SIGNAL TO NOISE AND SPECTRAL
RESOLUTION}
\footnote{ Based on observations with the NASA/ESA Hubble Space Telescope,
obtained
at the Space Telescope Science Institute, which is operated by the Association
of Universities for Research in Astronomy, Inc., under NASA contract
NAS5-26555}
\\[1.5cm]
{\em Ari Laor, John N. Bahcall, Buell T. Jannuzi, Donald P. Schneider}\\
{\normalsize Institute for Advanced Study, School of Natural Sciences\\
Olden Lane, Princeton, NJ 08540}\\[.4cm]
{\em Richard F. Green}\\
{\normalsize National Optical Astronomy Observatories\\
P.O Box 26732, Tucson, AZ 86726} \\[.4cm]
and \\[.4cm]
{\em George F. Hartig}\\
{\normalsize Space Telescope Science Institute\\
3700 San Martin Drive, Baltimore, MD 21218\\ [5cm]
Accepted by {\em The Astrophysical Journal}}\\
\newpage
\Large ABSTRACT \normalsize \ec
We analyze the ultraviolet (UV) emission line and continuum properties
of five low-redshift active galactic nuclei
(four luminous quasars: PKS~0405$-$123, H1821+643
 , PG~0953+414, and 3C273, and one bright Seyfert 1 galaxy: Mrk~205).
The HST spectra have higher signal-to-noise ratios (typically
$\sim 60$ per resolution element)
and spectral
resolution ($R = 1300$) than all previously-published UV
spectra used to study the emission characteristics of active galactic nuclei.
We include in the analysis ground-based optical spectra covering
\hb\ and the narrow [O~III]~$\lambda\lambda$4959,5007 doublet.

The following new results are obtained: \lyb/\lya=0.03$-$0.12 for the
four quasars, which is
the first accurate measurement of the long-predicted \lyb\ intensity in
QSOs. The cores of \lya\ and C~IV are symmetric to an accuracy
of better than 2.5\%  within about 2000~km~s$^{-1}$ of the line peak.
This high degree of symmetry of \lya\ argues against models in which
the broad line cloud velocity field has a significant radial component.
The observed smoothness of the \lya\ and C~IV line profiles requires at least
$\sim 10^4$ individual
clouds if bulk velocity is the only line-broadening mechanism.
The overall similarity of the \lya\ and C~IV~$\lambda$1549 profiles rules out
models for the broad line region (BLR) with a radial distribution of virialized
clouds having an ionization parameter $U\propto$ Radius$^{-1}$.
The measured high values of O~VI~$\lambda$1034/\lya\ and low values of
C~III~$\lambda$977/O~VI~$\lambda$1034
imply a BLR component with $U \sim1$. The excess red-wing flux in O~VI relative
to \lya\
suggests the presence of an inner, high-velocity, optically-thin
component with $U>1$ in the BLR.

The N~V/\lya\ ratio is $0.135 \pm 0.01$ for the four quasars, which may be
an indication of higher-than-solar N abundance and metallicity.
The maximum contribution of a narrow ( [O~III]-like) component is about $3-6$\%
of the total broad line flux; this limit is generally highest for C~III]. This
result constrains
the covering factor of the narrow line region or indicates the presence of
dust.
An unresolved component having full width at half maximum
$<230$~km~s$^{-1}$ typically
contributes less than 0.5\% of the observed broad lines flux.
The HST data permit the first relatively accurate measurements of the
\lyg, C~III~$\lambda$977, S~VI~$\lambda\lambda$933,945 , and
the N~III~$\lambda$991 emission lines, as
well as the measurement of a number of other weak or strongly blended
lines at $\lambda>1216$ \AA.

In agreement with observations of high-redshift quasars,
the peaks of \lya, C~IV, and C~III] are blueshifted by $\sim$
$200$~km~s$^{-1}$ relative to [O~III]~$\lambda$5007, while
He~II~$\lambda$1640 is shifted by
about $500$~km~s$^{-1}$. The low ionization lines, Mg~II, \hb, and
O~I~$\lambda$1304,  are in most cases only marginally shifted to the red.
\\[.3cm]
{\em Subject headings}: galaxies:nuclei---galaxies:Seyfert---
ultraviolet: spectra---line profiles---quasars
\bc \section{{\rm INTRODUCTION}}\ec

This paper analyzes the emission properties of
low-redshift active galactic nuclei (AGNs) observed
with the {\em Hubble Space Telescope} (HST) at a high signal-to-noise ratio
S/N ($\gtsimeq 60$ per resolution element) and a relatively high spectral
resolution ($\sim 230$~km~s$^{-1}$).
We study the spectra of four low-redshift quasars
($0.157\le z\le 0.573$):\ PKS~0405$-$123, H1821+643, PG~0953+414, and 3C273,
which have luminosities in the range $-28.4\le$ M$_{\rm V}\le -25.5$
(for H$_0$=50~km~s$^{-1}$, q$_0$=0) and Mrk~205, a bright Seyfert~1 galaxy
with M$_{\rm V}= -22.9$.
We use objective algorithms for modeling and deblending the emission line
profiles. The results presented here can help determine the
physical and dynamical conditions near the centers of AGNs.

The ultraviolet (UV) emission properties of low-redshift AGNs have been
 studied previously with the International Ultraviolet Explorer (IUE)
(e.g. Green {\it et al.} 1980; Ulrich {\it et al.} 1980; Kinney {\it et al.}
1985,
 1987). These
observations have a spectral resolution of 6~\AA, i.e. 1000~km~s$^{-1}$
at $\lambda = 1800$~\AA.
The small IUE mirror size allows high (S/N) observations
to be obtained only for the brightest
AGNs, and the observations are further limited by the presence of systematic
variations in
the detector calibration at a level of 5\% or more (e.g. Kinney {\it et al.}
1991).
High S/N observations at a higher spectral resolution for quasars are available
only
through ground-based observations of high-redshift objects. These high-redshift
studies are strongly
affected at $\lambda_{\rm rest}\le 1216$~\AA\ by absorption from the \lya\
forest systems.

The present analysis provides new information on the emission line regions of
low-redshift AGNs. In particular:

1. At $\lambda_{\rm rest}<1216$~\AA, we study low-redshift
AGNs without significant distortion by intervening absorption systems.
This allows us to make accurate measurements
of the degree of asymmetry of the \lya\ emission line, of the width of the O~VI
line, of the contribution of \lyb\ to the O~VI+\lyb\ blend,
and of the flux in C~III~$\lambda$977+\lyg\ and N~III~$\lambda$991.

2. At $\lambda_{\rm rest}>1216$~\AA, we study in detail
the profiles of the prominent UV emission lines and measure the
flux in very
weak UV emission lines. This was previously possible only using ground-based
observations of high-redshift quasars.

The outline of this paper is detailed below. In \S 2, we describe the
observations
 and data reduction, and in \S 3 we outline the line and continuum fitting
procedures. The results are presented in \S 4. In \S 5,
 we compare the line ratios we measure here with IUE observations of
low-redshift AGNs and with
ground-based observations of high-redshift quasars. In \S 6, we compare our
results
for the line profiles with previous studies and discuss the theoretical
implications. In \S 7, we compare our results for the line ratios with recent
theoretical predictions
 and discuss
some of the theoretical implications. We conclude in \S 8 with a summary of our
main results. The technical details of the line-fitting procedures are given in
the Appendix.
\newpage
\small
\bc \section {{\rm OBSERVATIONS}} \ec
\normalsize

\bc \subsection {{\em The HST Observations}} \ec

We observed five AGNs using the R=1300 gratings of the Faint Object
Spectrograph (FOS). Table 1 lists the five
objects in our
sample, together with their J2000 positions, redshift, $V$ and M$_{\rm V}$
magnitudes (calculated for H$_0$=50~km~s$^{-1}$, q$_0$=0), Galactic reddening,
and the papers that analyzed the absorption line
properties of these objects using HST data.
The coordinates were measured with the Space Telescope Science Institute's
Guide Star Selection System Astrometric Support Package and should be accurate
to $\simeq 1''$. The $V$ magnitudes are from the V\'{e}ron-Cetty \& V\'{e}ron
(1991)
catalog. The redshifts were measured using [O~III]~$\lambda$5007, as further
described below. The
reddening is deduced from the neutral hydrogen column density N$_{\rm H}$
(taken
from Stark et al. 1992), and the E(B$-$V) {\it vs.} N$_{\rm H}$ relation given
by Burstein \& Heiles
(1982). The alternative E(B$-$V) {\it vs.} N$_{\rm H}$ relation found by Savage
\& Mathis (1979) implies a significantly higher Galactic extinction, in
particular
for PKS~0405$-$123, where the Savage \& Mathis relation gives E(B$-$V)=0.078
instead of 0.031, and also for H1821+643,
where E(B$-$V)=0.083 instead of 0.035. Applying these higher, but equally
plausible,
reddening
corrections would result in a factor of about two higher flux at the shortest
UV wavelengths. More accurate N$_{\rm H}$ measurements for 3C273
and H1821+643 are given by Savage {\it et al.} (1993), who find N$_{\rm H}$
values higher than the Stark {\it et al.} values by 8\% and 4\%, respectively;
 these differences, however, are
negligible in comparison with the large uncertainty present in the  N$_{\rm H}$
to E(B$-$V) conversion relation mentioned above.

Table 2 lists the dates of the observations, the gratings and aperture used,
and exposure
times for each object. All observations were made through the
$0.25^{''}\times 2.0^{''}$ slit except for 3C273, where we have also added
exposures made through the 0.3$^{''}$,
0.5$^{''}$, and $1.0^{''}$ circular apertures.
The wavelength ranges covered by each grating are 1150~\AA$-$1606~\AA\ for the
G130H,
1600~\AA$-$2310~\AA\ for the G190H, and 2230~\AA$-$3280~\AA\ for the G270H.
The dispersions in the G130H, G190H, and G270H data are respectively 0.25,
0.36,
and 0.52~\AA\ pixel$^{-1}$, with approximate spectral resolutions (full width
at half
maximum) of 1.1~\AA, 1.5~\AA, and 2.0~\AA\ for the three gratings.
Further details concerning the instrumental configuration and the data
calibration
are described in Schneider {\it et al.} 1993, and in Bahcall et al. (1991,
1992a,b, 1993a,b). The long
exposure times combined with the high UV flux of all the objects
produced a S/N per resolution element in the range of $25-120$, with a typical
value of $\sim 60$. The spectrum of
PKS~0405$-$123 extends to the shortest rest wavelength ($\lambda_{\rm rest}
 \sim 730$~\AA). The spectrum of Mrk~205 extends to the longest rest
wavelength,
($\lambda_{\rm rest} \sim 3050$~\AA), but we have no FOS data for this object
for
 $\lambda_{\rm rest}<1500$~\AA. The data for 3C273 has a gap
at 1380~\AA\ $< \lambda_{\rm rest}< 1420$~\AA\ (Bahcall {\it et al.} 1991)
which makes
the O~IV]~$\lambda$1402+Si~IV~$\lambda$1397 blend unobservable.

\bc \subsection {{\em Ground-Based Observations}} \ec

We have obtained ground-based optical spectroscopy including the
H$\beta$~$\lambda$4861 + [O~III]~$\lambda$5007 lines for all objects. Most of
these spectra have
a similar resolution and S/N as the UV data. All objects, except
for PKS~0405$-$123, were observed at the Kitt Peak National Observatory 4-m
telescope.
The observations of 3C273 and PG~0953+414 are described in Boroson \& Green
(1992).
Mrk 205 was observed on 28 April, 1992, with the R-C spectrograph on
the Kitt Peak 4-meter telescopes.  The B\&L 400 grating and TI 5 CCD
gave coverage from 4900 to 7600~\AA\ at 12~\AA\ resolution.
The quasar H~1821+643 was observed on August 20th, 1992, with the Kitt
Peak 2.1-m telescope and Goldcam CCD spectrograph.  Grating 240, ruled
at 500 l/mm, gave coverage from 4400 to 9000~\AA, with 4.5~\AA\ resolution.
A high S/N optical spectrum of PKS~0405$-$123, extending from 3100 to
8700~\AA,
was kindly supplied  to us by B. Wills (published in Wills, Netzer \& Wills
1985).
The optical and UV flux densities of PKS~0405$-$123 in the
overlap region of the spectra ($\sim 3250$~\AA) agree to better than
6\%, even though these observations were made 9 years apart.
We were not able to intercalibrate the optical and UV spectra for the
other 4 objects due to lack of overlapping coverage.
The Fe~II optical blends were subtracted from the optical spectra of all
the objects using the template method applied by Boroson \& Green (1992)
for PG~0953+414 and 3C273.

The optical spectroscopy is used for the following purposes: 1.
To define the redshifts of all
objects in a uniform manner using the peak of the narrow [O~III]~$\lambda$5007
emission line. 2. To make a template  for the profile of the narrow emission
line  component using
the [O~III]~$\lambda$5007 line. 3. To compare the low-ionization
H$\beta$ profile to that of the high-ionization UV lines.

\bc \section {{\rm THE LINE AND CONTINUUM FITTING METHOD}} \ec

In order to extract the
information present in the emission line spectra, we have developed a specific
line-fitting method which has the following characteristics:
1. It implements an objective algorithm.
2. Spectral regions possibly affected by narrow absorption lines
can be identified and rejected from the fit.
3. It produces a smooth model for the intrinsic emission-line profiles and
allows an accurate measurement of the integrated emission fluxes.
4. The fit parameters can be physically interpreted as integrated flux,
velocity shift, and velocity dispersion.

Below we review briefly the details of the fitting method; complete details are
given in the Appendix. The first step is to define the continuum. A smooth
continuum
model, e.g. a single power-law, cannot follow the apparent continuum level to
better than
$\sim 10\%$, due to the presence of very broad and low level emission
features. We avoid modeling these features by making local continuum fits.
Each spectrum is divided
into sections that are about $\sim 200$~\AA\ wide, centered around the
prominent
emission features. For
each section the local continuum is defined as a power-law that intersects the
spectrum at the two end points. We model the observed emission line profiles as
a sum of Gaussians components and we use a $\chi^2$ minimization routine
to obtain the
best fit parameters. We found that at least three Gaussians were required to
obtain an acceptable model fit for the prominent UV line profiles
(\lya, C~IV, and C~III]), and
one Gaussian was usually sufficient for weak or strongly blended lines (e.g.
N~V). Each Gaussian is defined by three parameters, and the initial guess for
the
values of these parameters is obtained by measuring the zeroth, first, and
second
moments of the flux distribution within a given section of the emission line.
For the three Gaussians used to model the prominent lines, the initial guess
for
the fit parameters is obtained by measuring the moments
of the flux in the lower, middle, and upper section of the line. For the single
Gaussian component which models a blended line, we measure the flux moments
within 10 \AA\ of the rest wavelength of the line. The best fit values
obtained following the $\chi^2$ minimization are generally
not very far from the initial guess values, in particular when the profile is
characterized by a high S/N and is not significantly affected by absorption.

In some cases, it is possible to obtain an acceptable fit to the profile of
a blend using a sum of template line profiles. We constructed a symmetric
template profile
using the blue wing of \lya, and used it to deblend the O~VI, \lya, and C~IV
blends
in three of our five objects.
This method is used to test the significance of the apparent differences in the
line profiles, and also to get a better
estimate of the flux in some of the strongly blended lines (e.g. N~V, \lyb).
Further details are given in \S 4.2.5 and in the Appendix.

\bc \section {{\rm RESULTS}} \ec

\bc \subsection{{\em The Continua}} \ec

Figure 1 displays the spectra of the five objects.
The spectra are shown on rest wavelength scales; they are corrected for
Galactic
reddening using the extinction values given in Table 1 and the reddening law
of Seaton (1979). To improve the presentation, the displayed spectra were
smoothed
by median filtering over 21 pixels (except near the peaks of the prominent
lines).

The average spectrum of $\sim 700$ quasars observed from the ground by Francis
et al.
(1991, hereafter the Francis {\it et al.} composite) is also shown for
comparison in each panel. Note the overall
similarity of the continuum slopes at $\lambda\gtsimeq 1300$~\AA\ and of the
various emission features between each of our objects
and the Francis {\it et al.} composite, which represents significantly higher
redshift ($z \sim 1-2$) quasars. The large discrepancy shortward of Ly$\alpha$
is most
 probably due to
unresolved Ly$\alpha$ absorption systems which depress the continuum level in
the Francis et al. composite.
There is a broad emission feature in all spectra on the red side of
C~IV~$\lambda$1549 at
$\lambda\sim 1600$~\AA; this feature is particularly strong in H1821+643.

The spectral slope, $\alpha=d \ln F_{\nu}/d \ln \nu$, between adjacent
continuum windows is given for each object in Table 3. Note the large and
non-monotonic variations
of $\alpha$ with wavelength for each object. These variations indicate, as
mentioned in \S 3 and \S A.1, the presence
of broad quasi-continuum emission features in the spectra.

\bc \subsection{{\em The Lines}} \ec

Figures 2a-e show the spectra of each of the five objects in detail. The
spectra were
smoothed with a Gaussian as discussed in \S A.2. For the sake of clarity, all
points with $\Delta<-3.0$ were deleted from the plotted spectra, where $\Delta$
is
the deviation from the median in units of standard deviations (see \S A.2). The
spectra of all objects are plotted
in parallel to facilitate their intercomparison. We indicate the rest
wavelength
positions of most lines that have
been observed in the past or that were predicted to be present in AGNs. Most
lines are composed of several components, as indicated in each panel; the
effective
wavelengths of these
multiplets are taken from Morton (1991). A large number of
weak continuum features are apparent in each spectrum. Weak emission
features were measured only if they appear in more than one object and if they
have a plausible
identification (i.e. were clearly identified in other objects, or are expected
based on the presence of other emission lines). A large number of broad
emission features are present at $\lambda>2000$~\AA; most are likely to be
blends of Fe~II multiplets, as shown by Wills, Netzer \& Wills (1985).
The effects of blendings are minimized in narrow-line quasars. Therefore
the broad and blended emission features can be optimally studied by using a
narrow line quasar spectrum as a template, as done by Boroson \& Green (1992)
for the optical Fe~II emission blends; however, such a template is not yet
available in the UV.

\bc \subsubsection{{\em Line Deblending}} \ec

 The four prominent blends in the spectrum of each object are:
1. O~VI+\lyb+N~III+C~III+Ly$\gamma$,\\
2. \lya+N~V+O~I+C~II, 3. C~IV+He~II+O~III]+N~IV], and 4. C~III]+Si~III]+Al~III.
These lines were
deblended using the algorithm described in \S A.2. No FOS data at $\lambda_{\rm
rest}<1600$~\AA\ are available
for Mrk~205; for this object we fitted the C~IV and C~III] blends, and
Mg~II~$\lambda$2798.

Figures 3a-e display the sequence of fits that leads to the line deblending
results. The formal $\chi^2$ of the fit and the number of degrees of freedom
are
indicated in each panel. The value of $\chi^2$ only serves to indicate
the average level of deviation of the data from the best fit model, rather than
to measure the statistical significance of the fit (see \S A.2).
The continuum level is not well determined on the blue side of O~VI+\lyb\
due to the complicated continuum shape in that region.
In two objects, PG~0953+414 and 3C273, the continuum underlying O~VI+\lyb\
is set by extrapolation
of the slope measured at longer wavelengths. Since the observed continuum slope
in the UV tends to steepen towards shorter wavelengths (e.g. O'Brien,
Gondhalekar,
 \& Wilson 1988), this extrapolation might underestimate the true flux in the
blue wing of O~VI+\lyb, and produce the very asymmetric broad wings of
O~VI+\lyb\
in PG~0953+414 and 3C273 (Figs. 3.c,d).

Table 4 gives the best fit parameters for each of the prominent blends
displayed in Figs. 3.a-e, and a number of other lines/blends; the errors in the
fit parameters are also given. The fluxes given in Table 4 are in
the observed frame, and the values of the EW is given in the rest frame, i.e.
it is equal to the observed EW
divided by ($1+z$). Adding the superscript error of each parameter gives the
value
obtained for the fit with the continuum displaced upwards by 1$\sigma$, and
adding the subscript error gives
the value obtained with the continuum displaced downwards by 1$\sigma$ (see \S
A.2). As mentioned above, these two errors do not always have opposite signs.
The 9 parameters (of the 3 Gaussians) obtained
from the fit to the prominent lines can be useful when making
a statistical study of the line profiles of a larger sample of objects.

\bc \subsubsection{{\em The Apparent Line Asymmetry}} \ec

Figure 4 displays the asymmetry of the \lya\ and C~IV blends.
We show the
ratio of the red ($+v$) to blue ($-v$) wing flux in the best-fit model
as a function of
velocity from the line peak.
 The line peak is defined using a parabola fitted to the three highest
flux points in the best fit model.
We plot the ratio of the fit to the observed profiles, rather than
the deblended \lya\ and C~IV profiles,
since any deblending procedure is model dependent.
 In almost all cases both
\lya\ and C~IV  display a generally similar behavior of an increasing
red to blue wing flux ratio
with increasing velocity from the line center. \lya\ has a symmetric
core in two objects and C~IV in three objects. In both 3C273 and PG~0953+414
this ratio deviates by less than
5\% from unity out to a velocity of $\sim 2000-2500$~km~s$^{-1}$ from the line
peak.
 The C~IV profile is generally more symmetric than \lya. The symmetry of \lya\
is
further addressed below in \S 6.4.

Figure 4 also shows the expected asymmetry in the case of a cloud distribution
which is spherically symmetric in both position and velocities, and assuming
each
cloud emits isotropically. The asymmetry
in this case
results from special relativistic beaming which enhances the blue wing
emission.
In all cases, except C~IV in Mrk~205, the observed asymmetry has the opposite
sense.

\bc \subsubsection{{\em Comparison of the Line Profiles}} \ec

Figure 5 compares the fitted models for the profiles of the blends of
O~VI, \lya, C~IV, C~III],  \hb\ (with the Fe II subtracted), and the narrow
[O~III]~$\lambda$5007
line, for all objects. The profiles of all lines are scaled such that the
maximum flux density of the model fit is at 1.
The O~VI blend has the broadest profile in all four quasars, having an
 average FWHM$=5148\pm 432$~km~s$^{-1}$
(dispersion from mean, see specific values in Table 4), which is
about 50\% larger than the average FWHM$=3514\pm 569$~km~s$^{-1}$ of C~IV.
The FWHM of \lya\ is 3014$\pm 432$~km~s$^{-1}$, which is the smallest of the
broad
emission lines. The significance of these profile differences is addressed
below (\S 4.2.5)

In all objects, except H1821+643, C~III] has the narrowest peak.
In all cases, \lya\ is narrower than C~IV, except for the broad wings,
where it is comparable to C~IV. The broad wings of O~VI are significantly
asymmetric in 3C273 and in PG0953+414. This could be due to errors in the
extrapolated continuum
shape, as mentioned above (\S 4.2.1). The \hb\ line has a significant red wing
excess in PKS~0405$-$123 and in H1821+643, and its peak
in all objects appears to coincide well with the peak of [O~III] $\lambda$5007.
 The spectrum of H1821+643 is different from the spectra of the other four
objects in that all of its emission lines are strongly asymmetric,
 having a strong red excess.
The largest asymmetry in the spectrum of H1821+643 occurs in the \hb\ profile.
A similar asymmetry is apparent in the other Balmer
lines (Kolman {\it et al.} 1991).

Table 5 gives the velocity shifts of the lines. The observed wavelength of a
line
is defined by the position of the peak of the fitted model. In a few cases,
the observed line profile near the peak is asymmetric, and the fitted model
peak can
somewhat deviate from the observed
peak (most notably in the fit of O~VI in H1821+643, Fig.3b).  The velocity
shifts
are measured relative to the rest wavelengths of \lya, C~IV, C~III]
and O~VI (given in Table 4). All lines, except O~VI+\lyb\ in PG~0953+414
and 3C273, are shifted by less than 300~km~s$^{-1}$. The low ionization
lines, \hb\ and Mg~II, are generally only
slightly shifted, in most cases to the red, while the high ionization lines are
generally blueshifted. Note that the HST absolute wavelength calibration is
defined by a system in which the strong interstellar absorption lines in
the direction of each object are at rest, while the optical redshifts are
measured
with respect to the local standard of rest (LSR). This can introduce systematic
velocity shifts of up to $\sim 60$~km~s$^{-1}$ between the optical and UV
redshifts
(Bahcall {\it et al.} 1993b, Savage {\it et al.} 1993).

\bc \subsubsection{{\em The Narrow-Line contribution}} \ec

Table 6 presents the maximum possible contribution of a narrow [O~III]-like
component to
the total flux in the prominent UV lines and in \hb\ for each object. This
contribution is measured by repeating the fitting process described in \S A.2
with the further constraint that the velocity dispersion of the third
(narrowest) component be equal to that of [O~III] $\lambda$5007, as measured
for each
object. The FWHM of [O~III] is also given in Table 6. This added constraint
results
in a small increase in $\chi^2$ compared with the minimum value obtained
when all parameters
are allowed to vary. The largest contribution of a narrow-line-like
component generally occurs in \lya\ and C~III], and the lowest in O~VI.
The peaks of some of the emission lines are visibly affected by absorption
(e.g. \lya\ in PG0953+414). In these cases, the intrinsic line profile near the
peak is
probably narrower than
given in Table 6, and the maximum possible flux in a narrow component is
probably larger.

In order to estimate the maximum possible flux in a spectrally unresolved
component, we have repeated the fitting of the upper part of each line. We used
the
observed, rather than the Gaussian smoothed spectrum, and measured the flux of
the narrowest component which was forced to have a FWHM of 230~km~s$^{-1}$
(i.e. R=1300). We find the
upper limit on the flux in an unresolved component is generally less than 0.5\%
of the total line flux, and it is typically about 0.1$-0.2$\%. Only in one
case, the C~III] profile of PG~0953+414, is the unresolved component as large
as 1.3\%.
Thus practically all of the observed emission line flux is well resolved.

\bc \subsubsection{{\em Line-fitting with a template}} \ec

As mentioned above (\S 4.2.3), O~VI is broader (in FWHM) than \lya\ by $\sim
2100$~km~s$^{-1}$, on the average, and C~IV is broader than \lya\ by
500~km~s$^{-1}$.
However, it is not clear a priori whether the broader O~VI and C~IV profiles
reflect larger velocity dispersions, or just the fact that both these lines are
doublets. In the C~IV doublet (1548.20\AA, 1550.77\AA), the equivalent velocity
separation of of the two components is 499~km~s$^{-1}$, and in O~VI
(1031.93\AA, 1037.62\AA) it is 1651~km~s$^{-1}$. O~VI is also strongly
blended with \lyb\ (1025.72\AA) at a velocity separation of
2348~km~s$^{-1}$ from the mean O~VI wavelength.
In order to test whether C~IV and O~VI are intrinsically broader than \lya,
 we use the \lya\ profile as a template to fit the doublet lines. One cannot
just use the observed
\lya\ profile since its red wing is strongly blended with N~V. We therefore use
only
the blue wing of \lya\ to form a ``symmetric \lya'' template, and use this
symmetric profile to fit
the observed \lya, O~VI and C~IV blends (see \S A.4. for further
details on the fitting procedure).

Figure 6 presents the template fits. The $\chi^2$, the number of degrees of
freedom, and the velocity range over which the fit was made, are indicated in
each panel. We do not attempt to fit the blue wing of O~VI beyond
2000~km~s$^{-1}$, nor the blue wing of C~IV beyond 1000-2000~km~s$^{-1}$.
The best fit
parameters (flux and velocity shift) are given in Table 7. For each emission
line
we give two results, the upper row is obtained when we assume the ratio of
fluxes
of the components of a multiplet are the same as their statistical weight
ratio, and
the parameters in the lower row are obtained assuming equal weights for all
multiplet components. We arbitrarily fixed the velocity shifts of \lyb\ and
\lyg\ at the
best fit value of \lya, while for other weak components we generally assumed a
zero
velocity shift. We did not include
Mrk~205 in the fits since the \lya\ profile is not available for it; we also
do not include
H1821+643, since all its emission lines are strongly asymmetric and the
``symmetric \lya'' template cannot provide an acceptable fit. We attempted to
deblend the C~III] blend using the template method, however, in all cases
C~III] has
a significantly narrower peak than our template (see Fig.5), and an acceptable
fit
could not be obtained.

The core of the O~VI blend is well fit with the ``symmetric \lya'' template,
essentially all the difference in FWHM between \lya\ and O~VI can be explained
by
the wide separation of the O~VI doublet and some additional blending with
\lyb.  However,
in every case there is a significant excess in the red wing of O~VI compared
with the ``symmetric \lya'' best fit. The peak of
the O~VI blend is not well fit in PKS~0405$-$123 and PG~0953+414, and it is not
clear whether the effects of absorption can account for the magnitude of the
observed differences.

The red wing of C~IV cannot be fit by the template, possibly
 due to the $\sim 1600$ \AA\ emission feature (see \S 5.1.7). The peak of
 C~IV is narrower than the template, although \lya\ in the case of PG~0953+414
is
clearly self-absorbed, and the intrinsic \lya\ profile can have a stronger
narrow
peak than deduced here.

Figure 7 presents a template fit to the Si~IV+O~IV] blend. The Si~IV was
modeled as
a doublet at 1393.76 and 1402\AA, with a 2:1 flux ratio. The O~IV] multiplet
was modeled with 5 components at 1397.23, 1399.78, 1401.16, 1404.81, and
1407.38\AA,
and a 2:1:6:4:2 flux ratio (see Morton 1991). For each object we present
two possible fits with different Si~IV/O~IV] ratios, and assuming both
multiplets
are at rest with respect to [O~III]~$\lambda$5007. We do not measure the formal
best-fit ratio since the blend profile has a rather low
S/N, and the Si~IV/O~IV] ratio cannot be accurately determined.

\bc \section {{\rm COMPARISON WITH PREVIOUS OBSERVATIONS}} \ec

\bc \subsection{{\em Line Ratios}} \ec

We describe in this section the first measurements of a number of far UV
($\lambda<1000$~\AA) emission lines and many weak UV lines (such as
O~I~${\lambda}$1303, C~II~${\lambda}$1335,
O~III]~${\lambda}$1664, and N~III]~${\lambda}$1750) which were previously
measured only in high-redshift AGNs (e.g. Baldwin \& Netzer 1978; Uomoto 1984).

\bc \subsubsection{{\em S~VI~${\lambda}$937, C~III~${\lambda}$977+\lyg,
and N~III~${\lambda}$991}} \ec

The two components of the S~VI~${\lambda}$937 doublet (933.38, 944.52\AA),
are detected in the spectrum of PKS~0405$-$123 (Fig.2a). The S~VI emission has
not been previously detected
in the spectrum of an individual AGN. This line was detected in a
composite IUE spectrum of 22 low-redshift quasars derived by Sofia, Bruhweiler
\& Kafatos (1988), but the two components of the line were not resolved and
their flux was not measured. We note that
the S~VI~${\lambda}$937 doublet was also recently detected in absorption in the
broad absorption line quasar 0226~$-$1024 by Korista {\it et al.} (1993) in
data taken with the HST FOS.

We measure line flux ratios C~III+\lyg/\lya=0.056$-$0.064 in the spectrum of
PKS~0405$-$123 (Table 7) and 0.024 in the spectrum of H1821+643 (Table 4). The
C~III+\lyg\ blend has not been accurately measured before in quasars. The only
two papers we are aware of that measure this blend
in individual quasars are by Wilkes (1986), who found
C~III~$\lambda$977+\lyg/\lya=0.04 in one high-redshift radio selected quasar
(PKS~1614+051), and Green {\it et al.} (1980), who observed this line with the
IUE in
two quasars (PKS~1302~$-$102, and PG~1247~+268), and found
C~III~$\lambda$977+\lyg/\lya=0.084 and 0.105.
Note that the Green {\it et al.} ratios are likely to be significantly
biased towards high values since the low S/N of the IUE spectra allowed
detection
of the C~III~$\lambda$977+\lyg\ blend in only two of their five objects.
The C~III+\lyg\ blend was also recently measured in the Seyfert 2 galaxy
NGC~1068 by
Kriss {\it et al.} (1992) using the Hopkins UV Telescope (HUT), where a ratio
of C~III~${\lambda}$977+\lyg/\lya= 0.06 was obtained.

The C~III+\lyg\ blend in PKS~0405$-$123 is
visibly asymmetric (Figs.2a and 7), suggesting a significant \lyg\
contribution.
Our best fit template deblending (Table 7) suggests \lyg/C~III$\sim 1/3$,
or \lyg/\lya=0.017. The C~III~$\lambda$977+\lyg\ blend is also observed in the
spectrum of H1821+643; however, the spectrum in this object appears to have a
discontinuity at 973 \AA\ and no \lyg\ contribution is apparent.

We measure N~III~$\lambda$991/\lya\ ratios of 0.013, 0.014, and 0.0085, in
PKS~0405$-$123, H1821+643 and PG~0953+414.
The N~III~${\lambda}$991 line has not been previously detected in an individual
quasar
spectrum. It was detected but not measured in the IUE composite spectrum of
Sofia, Bruhweiler \& Kafatos (1988). This
line was recently detected in the Seyfert 2 galaxy NGC~1068 by
Kriss {\it et al.} (1992), who found N~III~$\lambda$991/\lya=0.031.

 A particularly large systematic error is possible in our measurements
of the fluxes in S~VI, N~III and C~III, due to the low EW of these lines,
the presence of
narrow absorption lines, and the uncertainty in the continuum placement at
$\lambda<1000$~\AA. In the analysis presented above and in the following
sections,
we use the line fluxes from Table 7 when available, rather than Table 4,  as
the Table 7 values are likely to be more realistic.

\bc \subsubsection{{\em O~VI$~{\lambda}$1034+\lyb}} \ec

We measured an average line flux ratio O~VI~${\lambda}$1034+\lyb/\lya+N~V$ =
0.34\pm 0.13$ (the error here and below is the dispersion about the mean).
The O~VI+\lyb\ blend has
been measured in intermediate redshift quasars using the IUE (e.g. Green {\it
et al.} 1980; Gondhalekar O'Brien \& Wilson 1986; Kinney {\it et al.} 1987) and
in high-redshift quasars using
ground-based observations (e.g. Baldwin \& Netzer 1978; Wilkes 1984, 1986;
Steidel \& Sargent 1987). In intermediate redshift quasars,
O~VI+\lyb/\lya+N~V$=0.24\pm 0.07$ [using the measurements
of 13 quasars with $z\sim 0.3-0.75$ made by Kinney {\it et al.} (1985, 1987)].
The IUE measurements are likely to miss some of the flux in
the broad wings of the O~VI+\lyb\ blend due to the typically very low S/N
of the IUE spectra. Our higher O~VI+\lyb/\lya+N~V ratio may therefore be more
realistic.

In radio selected high-redshift quasars
O~VI+\lyb/\lya+N~V$=0.18\pm 0.11$ (using the measurements of
Wilkes 1986). A very similar ratio, 0.17$\pm 0.09$, is obtained for
optically selected high-redshift quasars (Osmer \& Smith, 1976), and it
therefore
appears that our small sample of low-redshift quasars is significantly
different
from high-redshift quasars. However, Osmer \& Smith measured the flux only
within $\pm 5000$~km~s$^{-1}$ of the O~VI+\lyb\ blend
center, and they therefore excluded most of the flux in the broad component
of this blend.
Repeating our measurements, with the broad component of the
O~VI+\lyb\ blend excluded, we find O~VI+\lyb/\lya+N~V$=0.15\pm 0.05$, which is
consistent with the high-redshift quasars ratio.
Given the lower S/N and the large number of \lya\ absorption
systems which affect the high-redshift quasars spectra, one
cannot rule out the presence of a significant flux in a broad component
(FWHM$\sim 20,000$~km~s$^{-1}$) of O~VI in high-redshift quasars, as found here
in low-redshift quasars.
It is therefore possible that the apparent difference in the O~VI+\lyb/\lya+N~V
ratio between high and low-redshift quasars just reflects the difference in
data quality and  measurement methods, rather than being a real physical
difference.

We find \lyb/O~VI ratios of 0.25, $\sim 0.10$, 0.27, and 0.35 for
PKS~0405$-$123, H1821+643, PG~0953+414 and 3C273, using the template
deblendings (given in Table 7), where for H1821+643 we estimated
the \lyb\ fraction by following the template deblending procedure using the
observed
C~IV profile. These four values are likely to overestimate the intrinsic
\lyb/O~VI
ratio since the template fitting of O~VI does not include a significant amount
of
flux in the red wing of O~VI (Fig.6). When this flux is added (as calculated
from
the integrated O~VI blend flux given in Table 4), the revised values are:
0.16, $\sim 0.10$, 0.14, and 0.28. These ratios were obtained assuming the
2:1 multiplet ratio in O~VI, and a zero O~VI velocity shift. Somewhat lower
ratios
are obtained assuming a multiplet ratio of 1:1 in O~VI, in which case O~VI must
be
 blueshifted (see Table 7). Very few attempts have been made in the past to
deblend \lyb\ from O~VI.
Wilkes (1984) used the C~IV line profile to deblend \lyb\ from O~VI in two high
redshift quasars, and found \lyb/O~VI ratios of 0.31 and 0.46.

Using the template deblendings (Table 7) we find
\lyb/\lya=0.033, 0.030, 0.066, and 0.12 in PKS~0405$-$123, H1821+643,
PG~0953+414 and 3C273. Similar values,
\lyb/\lya=0.06, 0.10, were found in two high-redshift quasars by Wilkes (1984).

\bc \subsubsection{{\em \lya+N~V~$\lambda$1240+Si~II~$\lambda$1263}} \ec

The range of \lya\ rest-frame EW, 51~\AA$-136$~\AA, found here is typical
for the luminosity range of $-28.4\le M_V \le -25.5$ of our four quasars (e.g.
Netzer, Laor, \& Gondhalekar 1992). We therefore do not expect our objects to
be very
peculiar in terms of their emission line properties.

 A rather small dispersion is found in the N~V/\lya\ line ratios. We
 measure N~V/\lya=0.13, 0.12, 0.14, and 0.15 for PKS~0405$-$123,
H1821+643, PG~0953+414 and 3C273 (using Table 7 in all objects
except H1821+643, where we use Table 4).
Very little data is available on the N~V/\lya\ ratio in low-redshift AGNs,
since the
low quality of the IUE spectra did not allow N~V to be reliably deblended from
\lya. Green {\it et al.} (1980) found ratios of 0.04, 0.039, and 0.17 in the
IUE
spectra of PKS~0405$-$123, PG~0953+414, and PKS~1302$-$102.
Kinney {\it et al.} (1985) quotes an average
 value of 0.15 for 8 objects out 21, but they do not give further details.

In high redshift quasars N~V/\lya=0.32$\pm$0.13, using the Wilkes (1986)
measurements of N~V in 34 objects (see similar values in Osterbrock 1989;
Netzer 1990). The formal error in the mean value for high redshift quasars is
$\pm$0.02, and it therefore appears to deviate significantly from the value
found
here for low-redshift AGNs. It is, however, possible that the N~V/\lya\ ratio
in high-redshift quasars has been systematically overestimated. This can
result from an underestimate of the intrinsic flux in the \lya\ red wing,
which might happen if this estimate
is influenced by the shape of the \lya\ blue wing, which is significantly
absorbed in high-redshift quasars (see Wilkes 1986).

We find Si~II~${\lambda}$1263/\lya=0.026, 0.018, and 0.053 for PKS~0405$-$123,
PG~0953+414, and 3C273, using the template deblending scheme (Table 7, note
that the
detection in PG~0953+414 is very marginal).
Apart from 3C273, the Si~II~${\lambda}$1263 blend was not measured in other
low-redshift AGNs. We note, however, that this blend can be clearly discerned
in some of
the IUE spectra of low-redshift AGNs presented by Buson \& Ulrich (1990).
The Si~II~${\lambda}$1263 blend was measured in radio selected high-redshift
quasars by Wilkes (1986). Using her tabulated values we find
Si~II~${\lambda}$1263/\lya$=0.10\pm 0.05$, as measured for 19 objects. This
value
is, however, significantly biased since there are 31 more objects in the same
sample
where \lya\ was observed but Si~II~${\lambda}$1263 was too weak to be measured.
The high-redshift Si~II~${\lambda}$1263/\lya\ ratio is significantly larger
than our low-redshift measurements, but in view of the
 strong bias present in the high-redshift measurements, we conclude that there
is no well-established physical difference
in the Si~II~$\lambda$1263 emission between low and high-redshift quasars.

\bc \subsubsection{{\em O~I~${\lambda}$1304 and C~II~${\lambda}$1335}} \ec

We find O~I~${\lambda}$1304/\lya=0.022$-$0.034, and
C~II~${\lambda}$1335/\lya=0.004$-$0.038, using the line fluxes given in Table
4.
The O~I~${\lambda}$1304 and the C~II~${\lambda}$1335 lines are generally too
weak to be measured in individual objects using the IUE. However, these two
lines are
detected in a composite IUE spectrum of 27 Seyfert 1 galaxies made by
V\'{e}ron-Cetty, V\'{e}ron \& Tarenghi (1983),
from which we obtain ratios of O~I~${\lambda}$1304/\lya=0.013, and
C~II~${\lambda}$1335/\lya=0.0034. Ground-based spectroscopy of high-redshift
optically selected
quasars gives O~I~${\lambda}$1304/\lya=0.035$-$0.047 and
C~II~${\lambda}$1335/\lya=0.025$-$0.028, based on the
composites of Francis {\it et al.} (who include N~V with \lya) and Boyle
(1990).
We therefore conclude that C~II/\lya\ and O~I/\lya\ ratios measured here are
generally consistent with
the ratios found in both low-redshift Seyfert 1 galaxies and high-redshift
quasars.

\bc \subsubsection{{\em Si~IV~${\lambda}$1397+O~IV]~${\lambda}$1402}} \ec

The Si~IV~+O~IV] blend is detected in PKS~0405$-$123, H1821+643, and
PG~0953+414; for the three objects Si~IV~+O~IV]~/\lya=0.062, 0.068, and 0.086.
The high-redshift quasars composite spectra of Francis {\it et al.} and Boyle
(1990)
give  Si~IV~+O~IV]~/\lya=0.19 and 0.10 respectively. There is no complete study
available of the Si~IV~+O~IV] blend in low-redshifts AGNs as it is rather weak
for IUE observations.
The only measurements we found are by Tinggui, Clavel \& Wamsteker (1992) who
analyzed IUE spectra of low-redshift
AGNs which were selected to have strong Si~IV~+O~IV] emission. Using their
tabulated values for 11 objects we find Si~IV~+O~IV]~/\lya$=0.18\pm 0.06$.
Our values are, as expected, significantly smaller
 than the measurements of Tinggui {\it et al.}, and appear to be
consistent with the ratio found by Boyle (1990) for high redshift quasars.

The Si~IV~${\lambda}$1397 and O~IV]~${\lambda}$1402 lines are strongly blended.
We have attempted to make a direct deblending of the
lines in PKS~0405$-$123 and PG~0953+414, where the template method could
be applied (\S 4.2.5). Although our S/N are relatively high we could not make
an accurate
determination of the line ratios; we find, however, that a ratio of
Si~IV/O~IV]$\sim 1-3$ is consistent with our data (Fig.7).
Baldwin \& Netzer (1978) have deblended Si~IV and O~IV] in 13 high-redshift
quasars
using a single template for each line and found that the blend is generally
strongly
dominated by O~IV]. An alternative common technique to estimate the relative
contributions of Si~IV and O~IV] is to use the mean wavelength of the blend.
Using this approach Wills \& Netzer (1979) found S~IV/O~IV]$\sim 0.18$.
However,
in a recent analysis using the same methods, but with more detailed modeling of
the expected mean wavelength, Tytler \& Fan (1992) found Si~IV/O~IV]$\sim 1$.

Both the SI~IV and the O~IV] multiplets have components with a rather large
wavelength separation and different flux ratios; as a result the line profiles
we constructed using the symmetric template
are quite asymmetric. For example, a 1:1 Si~IV/O~IV] ratio results in a very
asymmetric blend
profile having a significantly stronger red wing (Fig.7). Deblending both lines
using an identical symmetric template would lead in this case to a
significant overestimate of the O~IV] contribution. This might explain the
significantly lower Si~IV/O~IV] ratio found by Baldwin \& Netzer (1978), and
implies that the Si~IV/O~IV] ratio found here might also characterize
high-redshift
quasars.

Our estimates of the Si~IV/O~IV] ratio were done assuming both lines are not
systematically shifted from rest. If both lines are significantly blueshifted,
then the implied Si~IV/O~IV] ratio would be lower. We cannot
rule out a contribution of S~IV~$\lambda$1410 of the order of 10\%. This
contribution will be larger if Si~IV and O~IV] are significantly blueshifted.

\bc \subsubsection{{\em C~IV~$\lambda$1549+N~IV]~$\lambda$1486+
Si~II~$\lambda$1531}} \ec

Using the template deblendings we find a remarkably small dispersion in the
C~IV/\lya\
line ratios. Specifically, we find C~IV/\lya=0.47, 0.48, and 0.46 for
PKS~0405$-$123, PG~0953+414 and 3C273. We estimate that this ratio is 0.6 in
H1821+643
(where we include only half the flux of the broad C~IV component, as we assume
the
other half is due to the $\sim\lambda$1600 feature). The C~IV/\lya\ ratio has
already been determined for a large number of
AGNs. Using the measurements of Kinney {\it et al.} (1985, 1987), we find
C~IV/\lya=0.47$\pm0.16$ averaged over 14 AGNs with $z\sim 0.3-0.75$. This
suggests that our small sample has emission line ratios which are
typical of low to intermediate-redshift AGNs.

The N~IV]~${\lambda}$1486/\lya\ ratios we find are 0, 0.009, 0.008, and 0.025
for PKS~0405$-$123, H1821+643, PG~0953+414 and 3C273.
The N~IV]~${\lambda}$1486 line has not been previously measured in low-redshift
AGNs.
The high-redshift composite spectra of Boyle (1990) and Cristiani \& Vio
(1990) give N~IV]/\lya=0.01$-$0.03. The N~IV]/\lya\ ratios found here are
consistent
with the range of average values found in high-redshift quasars.

We measured Si~II~$\lambda$1531/\lya=0.038,
0, and 0.047 in PKS~0405$-$123, PG~0953+414 and 3C273, using the ``symmetric
\lya''
template fitting. However, the Si~II~$\lambda$1531 doublet is very strongly
blended with C~IV, and the measured values are based on the
assumption that the intrinsic profile of C~IV is identical
to the profile of \lya. If C~IV is intrinsically broader than \lya\ by
$\sim 10\%$, than the
flux in Si~II~$\lambda$1531 will be consistent with zero for all objects.
The Si~II~$\lambda$1531 doublet was previously measured by Ulrich {\it et al.}
(1980) in 3C273 with a flux similar to the one found here. Four more detection
in Seyfert galaxies are cited in Dumont \& Mathez (1981), and no reports of
this feature in high-redshift quasars were found in the literature.

\bc \subsubsection{{\em He~II~${\lambda}$1640+
O~III]~${\lambda}$1664 and the $\sim\lambda$1600 feature}} \ec

We find He~II~$\lambda$1640/C~IV=0.036, 0.021, 0.065, 0.022 and 0.10
and O~III]~$\lambda$1664/C~IV=0.073, 0.060, 0.032, 0.060, 0.061
in PKS~0405$-$123, H1821+643, and PG~0953+414, 3C273, and Mrk~205.
However, He~II is blended with the unidentified $\sim 1600$~\AA\ feature; this
probably introduces a large and nonuniform systematic error in the measured
He~II flux.
The O~III] line is probably less affected by blending, as also suggested by the
smaller dispersion in the O~III]/C~IV ratios.
We are not aware of any previous measurements of these lines in low-redshift
AGNs
(excluding 3C273).
In high-redshift quasars, Boyle (1990) and Uomoto (1984) find
He~II/C~IV$=0.12-0.13$ which is larger than our values,
and O~III]/C~IV$=0.08-0.10$, which is consistent with the ratios found here for
low-redshifts AGNs.
The He~II/C~IV measurements in high-redshift quasars are probably biased
towards
high
values since they do not allow for blending with an underlying component, as
done here. However, since He~II cannot be reliably deblended,
it is not clear a priori which of the He~II/C~IV measurements is likely to be
more realistic.

The unidentified broad emission feature at $\sim 1600$~\AA\ was already noted
in earlier studies, e.g. Wilkes (1984) and Boyle (1990).
We did not use a separate Gaussian component to model this
broad feature, since the
three Gaussian components used to fit the C~IV profile generally also allowed a
satisfactory fit to this feature. The $\sim 1600$~\AA\ feature mainly affected
the parameters of the broad Gaussian
component that describes the base of C~IV, resulting in a shift of its center
by $\sim 1700-8200$ km~s$^{-1}$,
and a very large FWHM $\sim 12,000-24,000$ km~s$^{-1}$.
The strength of the $\sim 1600$~\AA\ feature varies considerably among the five
objects in our
sample. Characterizing the strength of the $\sim 1600$~\AA\ feature by the
ratio
$F_{\lambda}^{\rm ob}/F_{\lambda}^{\rm c}$ (see \S A.2) at $\lambda=1610$~\AA,
we get a maximum
value of 1.32 in H1821+643, and a minimum of 1.08 in Mrk~205.

The $\sim\lambda$1600 feature is probably not an extended red wing of
C~IV because of the large velocity separation ($\sim 10,000$~km~s$^{-1}$)
from the peak of C~IV.
 Wills, Netzer \& Wills (1980) suggested that low level Fe II emission is
present
in the $\sim 1610-1680$~\AA\ range.

\bc \subsubsection{{\em N~III]~${\lambda}$1750 and Si~II~${\lambda}$1814}} \ec

The N~III]~${\lambda}$1750/C~IV ratio found here is $0.0-0.016$ {\it vs.} 0.07
in higher redshift quasars (Boyle 1990;
Uomoto 1984), while the Seyfert composite of V\'{e}ron {\it et al.} (1983)
gives
N~III]/C~IV$\sim 0.03$. The N~III] line is very weak in our spectra and
therefore
has a large measurement error. Measurement uncertainties are not
likely to be large enough to explain the
factor of $\sim 5$ difference between our results and these for higher redshift
quasars.

 We find Si~II~${\lambda}$1814/C~IV$=0.0-0.01$, with about a factor of two
possible error in this ratio due to the very low amplitude of the
Si~II~${\lambda}$1814 triplet.
We note, however, that some Fe~II emission is possible at 1820$-$1840 \AA\ (see
Wills {\it et al.} 1980); Fe~II emission might strongly contaminate, or
possibly
dominate, the flux attributed here to Si~II~${\lambda}$1814.
The Si~II~${\lambda}$1814 triplet was previously suggested in one Seyfert
galaxy
(private communication cited by Dumont \& Mathez 1981). This feature has not
been
detected in high-redshift quasars.

\bc \subsubsection{{\em Al~III~$\lambda$1857 + Si~III]~$\lambda$1892
+ C~III]~$\lambda$1909}} \ec

The C~III]~$\lambda$1909 line is strongly blended with the
Si~III]~$\lambda$1892 line, and somewhat
blended with the Al~III~$\lambda$1857 doublet.
We find a ratio of Al~III]/C~III]=0.05$-$0.33 with a mean value of 0.16,
which is consistent with the ratios of 0.14 and 0.12 obtained for
high-redshift quasars
by Gaskell, Shields \& Wampler (1981) and by Steidel \& Sargent (1991).
We note, however, that two Fe~II multiplets are expected in the 1849$-$1868\AA\
range
(Wills {\it et al.} 1980), and these might dominate the flux we attribute to
the
Al~III doublet. Both Si~III]~$\lambda$1892 and Al~III~$\lambda$1857 have not
been previously detected in the spectra of low-redshift AGNs (except 3C273).

We measure Si~III]/C~III]=0.03$-$0.35 with a mean value of 0.15, which
is roughly consistent with the value of 0.22 obtained by Gaskell {\it et al.}
for high redshift quasars.
However, Steidel \& Sargent find the Si~III] contribution to be consistent
with zero in $\sim 90$\% of their 92 objects. It is not clear why the large and
heterogeneous sample of Steidel \& Sargent differs from our sample and from
the sample of Gaskell {\it et al}.

\bc \subsection{{\em Earlier observations of our five AGNs}} \ec

All five AGNs described in this paper were previously observed with the IUE.
The observations of PKS~0405$-$123 and PG~0953+414 are
described in Green {\it et al.} (1980). A detailed analysis of the observations
of 3C273 is described in Ulrich
{\it et al.} (1980). Measurements of UV lines in Mrk~205 are given by Buson \&
Ulrich (1990), and the observations of H1821+643 are described
in Kolman {\it et al.} (1991).

A number of the emission lines
measured here were not measured or detected in the IUE spectra of our five
AGNs.
In particular,
most lines that are weaker than the Si~IV+O~IV] blend could not be reliably
measured in the IUE spectra. A comparison of
the line fluxes measured with both the HST and the IUE indicates
differences ranging from $\sim 10$\% to about a factor of two. A partial
explanation for
the differences is intrinsic variability of the emission line fluxes
during the typically 5-10 year interval between the IUE and the HST
observations.
 In some lines the different fluxes are probably also related
to the different wavelengths at which the continuum windows are set.
 For example, in C~IV we use a continuum window at 1720~\AA,
while previous studies usually employed a window at $\sim 1600$~\AA, which can
be strongly affected by the unidentified $ \sim 1600$~\AA\ feature (see \S
5.1.7).
In the case of weaker lines, such as the Si~IV+O~IV] blend, it is likely
that some of the difference results from the relatively low S/N of the IUE
measurements. In the case of 3C273, there are two very weak lines,
Si~II~${\lambda}$1194, and [Ne~V]~${\lambda}$1575, which Ulrich {et al.} (1980)
identified
in the IUE spectra, but which we do not detect in our HST spectra. The
significantly higher S/N and spectral resolution of the HST data allows
us to place an upper limit on the flux in these lines which is about three
times
lower than the flux measured by Ulrich {\it et al.} (1980). We confirm the
Ulrich {\it et al.} detection in 3C273 of the Si~II~$\lambda1263$ multiplet,
and we possibly also detect the Si~II~$\lambda1531$ multiplet (see below),
although
the fluxes we measure for the two Si~II lines differ by $\sim 50$\% from
the IUE values.

\bc \section{{\rm PHYSICAL IMPLICATIONS OF THE LINE PROFILES}} \ec

\bc \subsection{{\em The Number of Clouds in the Broad-Line-Region}} \ec

If the profiles of the line emitted by each individual cloud in the
broad-line-region (BLR) is
unresolved (i.e. narrower than about 230~km~s$^{-1}$), and the number of
individual
clouds contributing to the integrated line flux is significantly smaller
than the number of observed photons, then the fluctuations in the number of
clouds contributing to each spectral resolution element will result in a
statistically significant structure in the observed line profile.
Our fits to the line profiles (Figs. 3a-e) are generally characterized by
 $\chi^2_r\sim 1-2$. Given the high S/N of our spectra, this low $\chi^2$
implies
a remarkable lack of small scale structure in the line profiles over that
predicted from the finite number of photons observed.
Using the strongest UV lines, we find that the number of clouds is typically
larger than a few times $10^4$. This argument gives a rough lower
limit on the number of clouds, since some of the apparent deviations from the
smooth model fit could
be due to uncalibrated low level structure in the detector response, or weak
absorption lines. A similar
lower limit on the number of clouds was obtained by Capriotti, Foltz \& Byard
(1981)
and by Atwood, Baldwin \& Carswell (1982) based on the smoothness of the \ha\
and
\hb\ profile.

The limit obtained above depends
on the width of the line emitted by each individual cloud.
Significant broadening, for example by electron scattering (see
Shields \& McKee 1981; Emmering, Blandford \& Shlosman 1992 ), can reduce by
an arbitrary amount the lower limit on the number of clouds in the BLR.

\bc \subsection{{\em Line Velocity Shifts}} \ec

As first noticed by Gaskell (1982), different emission lines yield
systematically
different values for the redshift. A recent detailed analysis of systematic
velocity
shifts is presented by Tytler \& Fan (1992).

We find (see Tables 5 and 7) that
 the peaks of \lya, C~IV, and C~III] are typically blueshifted by
$150$~km~s$^{-1}$ to $250$~km~s$^{-1}$ (relative to [O~III]~$\lambda$5007),
while the low
ionization lines Mg~II, and \hb, are in most cases only marginally shifted
to the red. For He~II~${\lambda}$1640 we find
the largest average blueshift, $-476\pm 98$~km~s$^{-1}$ (excluding 3C273 with a
shift of $-$983~km~s$^{-1}$) with respect to
[O~III]~${\lambda}$5007, or $-$507~km~s$^{-1}$ with respect to H$\beta$.
(The possible systematic error is about 60~km~s$^{-1}$, see \S 4.2.3).

The template fits for O~VI (using the statistical-weights multiplets ratios)
are consistent with a zero velocity shift (Table 7),
although the Gaussian decomposition (Table 5) appeared to suggest large shifts.
This is partly caused by the relatively low S/N of the O~VI profile.
Another source for a large systematic error in the shift concerns the
estimate of the mean wavelength
of the O~VI doublet. The value used for the Gaussian decomposition is
1033.816 \AA\
(Morton 1991), which is the statistical-weight mean of the two doublet
components.
However, due to the large velocity separation of the doublet components
(1651~km~s$^{-1}$), the observed peak is dominated by the stronger
component which is at -548~km~s$^{-1}$ relative to the mean wavelength. Thus,
as clearly
seen in Fig.6, the peak of the fitted O~VI profile appears to be blueshifted,
even though the fit was made using O~VI at rest. The shift in the position of
the peak depends
on the amount of blending of the doublet components, i.e. the ratio of the
template width
to the doublet velocity separation. Another important source for a systematic
error
in measuring velocity shifts is the ratio of fluxes in the multiplet
components.
The weighted
mean multiplet wavelengths used here are appropriate for the case of
absorption.
In the case of emission, the ratio of fluxes in a multiplet can deviate from
the ratio of statistical weights (e.g. Netzer \& Wills 1983). If
all components are optically
thick, thermalized, and the emitting gas is isothermal, then the flux ratio
is practically one. As shown in Table 7, template deblending with a flux ratio
of one implies higher blueshifts, in particular for O~VI where the velocity
separation of the doublet is largest.

The relatively small velocity shifts found here for O~VI, \lya, C~IV, and
C~III],
are consistent with the results of Tytler \& Fan,
which are based on ground-based spectra of high-redshift quasars. Our results
differ from those obtained in earlier analysis of high-redshift quasars where
larger velocity shifts were found (e.g. Corbin 1990).

Tytler \& Fan find a He~II velocity shift of $-$454~km~s$^{-1}$
using the Balmer lines, rather than [O~III], to determine rest frame redshift.
This value is very similar to the $-$507~km~s$^{-1}$ average shift found here.
 For O~I~${\lambda}$1303, we measure an average shift of
$261\pm 129$~km~s$^{-1}$ relative to H$\beta$,
while Tytler \& Fan find $-$50~km~s$^{-1}$. One trivial source for the
difference
is the fact that we use O~I rest wavelength of 1303.49~\AA\ (Morton 1991), {\it
vs.}
1304.46~\AA\ used by Tytler \& Fan. Using the Tytler \& Fan value for the rest
wavelength reduces our shift to $88\pm 129$~km~s$^{-1}$, which is consistent
with
their result. Another complicating effect in the case of O~I is
the possible contribution of Si~II~${\lambda}$1308 (Fig.2b). Blending with
Si~II~${\lambda}$1308 could result in the apparent redshift of the O~I line
found
here and in earlier studies of high redshift quasars (e.g. Wilkes 1984).

  We find that the other UV lines are generally either too weak or too blended
for a
reliable determination of the position of their peak.

\bc \subsection{{\em A Comparison of the Line Profiles}} \ec

As shown in Fig.5, in all four quasars \lya\ has a narrower peak than C~IV, but
similar
broad wings (average FWHM=3514~km~s$^{-1}$ for C~IV vs. 3014~km~s$^{-1}$ for
\lya).
A similar effect was noted by Wilkes (1984) in high-redshift quasars, and by
Buson \& Ulrich (1990) in low-redshift AGNs. Some of this difference results
from
the fact that C~IV is a doublet with a velocity separation of 499~km~s$^{-1}$,
and, as shown in Fig.6, one can get a reasonable fit to C~IV using the \lya\
template. However, the fit in Fig.6 assumes some contribution from the
Si~II$\lambda$1531 doublet, and the agreement is not good in the red wing
of C~IV beyond $\sim 1000$~km~s$^{-1}$, possibly due to blending with the
$\sim 1600\lambda$ feature. Given these
possible blending effects we cannot reliably determine the reality of the
apparent
small differences between the \lya\ and C~IV profiles (see Fig.5).

The C~III] line generally has the narrowest peak of the broad lines, despite
the fact it is blended with
Si~III] and Al~III (and possibly also Fe~II). The relative intensity of
C~III] is suppressed at densities above a few times $10^9$~cm$^{-3}$.
The narrow peak of C~III] therefore suggests that the low velocity gas,
 possibly at larger distances, has a lower density and is a more efficient
emitter
of C~III] than the high velocity gas.

Detailed theoretical predictions of the relative line
profiles were made by Rees, Netzer \& Ferland (1989), who calculated the line
emission for various spherically symmetric
distributions of mass-conserving clouds and various scalings of the ionization
parameter $U$ (= density of ionizing photons/gas density), with radius.
The motion of the clouds was assumed to be virialized, i.e. line width $\propto
R^{-1/2}$, where $R$ is the distance of the
cloud from the central continuum source. Our results rule out their model
with $U\propto R^{-1}$, as it predicts differences in the line profiles which
are
significantly larger than observed here. Our results are consistent with the
Rees
{\it et al.} model with $U=$constant, which predicts only small differences in
the line profiles and that C~III] should be narrowest. The Rees {\it et al.}
model with
$U\propto R^{-1/2}$ predicts larger differences, but it cannot be clearly
ruled out by our data.

Fig.5 shows that the O~VI blend is significantly broader than all other strong
emission lines (average FWHM=5148~km~s$^{-1}$, {\it vs.}
3014~km~s$^{-1}$ for \lya, see \S 4.2.3).
However, the template fitting (Fig.6) indicates that despite this large
apparent
difference the O~VI blend can be well fit
(excluding the red wing at $v>2000$~km~s$^{-1}$) with a
sum of symmetric \lya\ components, and
the larger apparent width of the O~VI blend can be all attributed to the
large velocity
separation of the O~VI doublet (1651~km~s$^{-1}$) and the additional
blending with \lyb\ (at $-$2348~km~s$^{-1}$). The width of the O~VI blend,
relative
to the width of its
individual components, depends on the ratio of
the individual component width to the velocity separation of the doublet. This
broadening effect will be largest in objects with the narrowest lines, and we
generally expect O~VI will have a systematically larger width
than all other lines.
The claim made by Osmer \& Smith (1976) that the widths of O~VI and \lya\ in
high-redshift quasars are comparable must be due to the combination of
intervening absorption systems and low S/N spectra.

As shown in Fig.6, the red wing of O~VI ($v>2000$~km~s$^{-1}$) is not well fit
by
the \lya\ template. The excess flux in the observed broad red wing is
$\sim 20-40$\% of the flux fit with the \lya\ template. If the observed line
width
is due to bulk velocity of the emitting gas, then it implies that the high
velocity
gas produces emission with a significantly higher O~VI/\lya\ ratio than the low
velocity gas. According to the `standard' BLR photoionization models
(e.g. Kwan \& Krolik 1981; Rees, Netzer \& Ferland 1989) the relative flux
in O~VI increases with increasing ionization parameter. In particular, Ferland
{\it et al.} (1992) find that O~VI becomes the strongest line once
the ionization parameter $U$ is larger than 0.5
(assuming a gas density $n=10^{10}-10^{11}$ cm $^{-3}$).
This suggests that the high velocity gas responsible for the red wing of O~VI
is subject to a high ionization parameter. If the gas velocity is determined
by the depth of the potential well, then this implies the presence of clouds
with  $U$  significantly larger than 1 at small distances from the ionizing
continuum source.

There are a number of Fe~II multiplets between
O~VI and \lya. No reliable prediction of their strength is available.
It is possible that the apparent excess flux in the wing of O~VI
discussed above is mostly due to Fe~II emission.

The peak of the low ionization H$\beta$ line is systematically redshifted with
respect to the other prominent UV lines. It also appears to have a distinctly
asymmetric shape in two of our objects, where the red wing shows excess
emission. Some of this excess might be due to residual Fe~II emission, since,
as noted by Boroson \& Green, the fixed Fe~II template used here (based on
I~Zw~1)
is not a good representation of the Fe~II emission in some objects.
 However, residual Fe~II is not likely to be responsible for the strong
line asymmetry in H1821+643, since all other lines in this object display the
same
characteristic asymmetry.

The clearest difference between the low and high
ionization lines is that the later are typically
blueshifted ($\sim 150-250$~km~s$^{-1}$, and $\sim 500$~km~s$^{-1}$ for
He~II~$\lambda 1640$) relative to the low ionization
lines. The model suggested by Collin-Souffrin {\it et al.} (1988, see also
Dumont \& Collin-Souffrin 1990, and references therein), invokes
separate populations of high and low column density clouds having
different spatial configurations, and presumably different kinematics,
which are subject to different ionizing continua. It is not clear
whether the overall similarity of the low and high ionization line profiles
is consistent with such a model.

\bc \subsection{{\em Symmetry of the Line Profiles}} \ec

\bc \subsubsection{{\em Measured Limits on the Asymmetry}} \ec

The observed \lya\ blend is highly symmetric in  PG~0953+414, and 3C273
(see Fig.4). We find $\left| F_{\lambda (v)}/F_{\lambda (-v)}-1 \right| \le
0.04$ for
$0\le v\le$ 2000~km~s$^{-1}$, where $v$ is the velocity shift
(positive for redshift) from the peak of the best-fit model to the observed
line profile,
and $F_{\lambda (v)}$ is the flux of the best-fit model.
The template fitting of \lya+N~V (Fig.6)
indicates that no intrinsic asymmetry of \lya\ is required even at
velocities significantly above 2000~km~s$^{-1}$. In three of the four quasars,
PKS~0405$-$123, PG~0953+414, and 3C273, practically all of the observed
red-excess of \lya\ can be
attributed to blending with a N~V doublet whose two components have the
same symmetric profile assumed for \lya\ (the ``symmetric \lya'' template).
The high degree of symmetry of \lya\ in a significant fraction of AGNs was
noted
previously by many authors. In particular, in low-redshift AGNs by
Buson \& Ulrich (1990), in intermediate redshift AGNs by Kinney {\it et al.}
(1987), and in high-redshift AGNs by Wilkes \& Carswell (1982),
but no quantitative limit was put on the possible asymmetry of \lya\
in these studies.

\lya\ is predicted to be blended with the semi-forbidden O~V]~$\lambda$1218
(see Baldwin \& Netzer 1978). This line can be used as a tracer of gas with
a high density and ionization parameter. In particular,
Ferland {\it et al.} (1992) predict O~V]/\lya $>0.1$ for $U>0.1$ and
$n=10^{11}$~cm$^{-3}$.
Given the assumption that \lya\ is intrinsically symmetric we can rule out an
O~V]/\lya\ ratio larger than 3-10\%.

The observed C~IV profile also has a very symmetric core, with
$ \left| F_{\lambda (v)}/F_{\lambda (-v)}-1 \right| \le 0.05$ for $0\le v\le$
2000~km~s$^{-1}$,
 in PKS~0405$-$123, PG~0953+414, and 3C273. The observed red excess asymmetry
in C~IV might be related to blending
with the $\lambda\sim 1600$~\AA\ feature. Buson \& Ulrich (1990)
found C~IV to be very symmetric in IUE spectra of low-redshift AGNs.
However, Young {\it et al.} (1982), Wilkes (1984, see detailed profiles in
Kallman {\it et al.} 1993), and Ulrich (1989) generally find
a blue, rather than red, excess asymmetry in C~IV in high-redshift quasars.
Some of
the difference might result from their use of a $\sim 1600$ \AA\ continuum
window,
rather than the $\sim 1700$ \AA\ window (used here) which generally has a lower
flux, and therefore implies more flux in the red wing of C~IV.

The symmetry of C~III] could not be addressed reliably due to the strong
blending
with the Si~III] line.

\bc \subsubsection{{\em Constraints on the Models}} \ec

The very high degree of symmetry of \lya\ presents a strong constraint on
models for the cloud dynamics in the BLR.
\lya, and possibly also C~IV for high values of $U$, are expected to be emitted
highly anisotropically from each of the broad line clouds
(Ferland, Netzer \& Shields 1979, Ferland {\it et al.} 1992). Thus models
invoking a pure radial cloud velocity field
will produce
a strong line asymmetry, and such models are ruled out. A similar
conclusion was recently reached by Kallman {\it et al.} (1993), based on a
comparison of the observed prominent line profiles with predictions of models
which combine cloud kinematics and photoionization calculations.
Some of this asymmetry might be lowered by preferential obscuration
of part of the BLR, or by electron scattering of the ionizing continuum
(Kallman \& Krolik 1986; see also some examples in Netzer 1990). It seems
unlikely that the combination of these effects will balance the line
asymmetry to the low level measured here. A radial
velocity field for the clouds in the BLR has also been ruled out in a few AGNs
based on the response of the emission line profiles to continuum variations
(e.g. Maoz {\it et al.} 1991; Koratkar \& Gaskell 1991).
Models involving an isotropic cloud velocity
distribution at each distance produce more symmetric profiles (note that, as
shown in Fig.4,
special relativistic effects still produce a noticeable asymmetry even in this
case), but these
models also imply that all the clouds in the BLR collide with each other within
a few dynamical time scales, i.e. a few dozens of years. Alternatively, the
observed line width might not be related to bulk motion but rather
reflect broadening produced by optically thick
inelastic electron scattering. This requires a rather high column density of
obscuring warm gas in the BLR (cf. Shields \& McKee 1981; Emmering, Blandford
\&
Shlosman 1992).

\bc \subsection{{\em Narrow Line Contribution}} \ec

We find that the C~III] line has the highest average narrow-[O~III]-like
line component,
followed by \lya, with O~VI having the smallest narrow line contribution. The
contribution of the narrow component does not appear to be related to the
objects'
luminosity in our small sample. The rather low narrow line contribution to
Mg~II found here was previously noted in high-redshift quasars by
Grandi \& Phillips (1979). However, the peaks of Mg~II, C~IV, and in particular
O~IV, are broadened as they are unresolved doublets (velocity-equivalent
separations are respectively 770, 500, and 1650~km~s$^{-1}$), and therefore the
maximum contribution of the narrow component to these doublet cannot be
reliably
measured without more detailed modeling.

The maximum narrow-line-like component of the broad permitted lines was
recently
studied in detail by Wills {\it et al.} (1993) in a sample of 7 radio-loud
quasars.
They found that the upper limits on the narrow-line contributions to the
prominent UV lines
are significantly smaller than the upper limit in \hb.
However, we find the upper limit on the narrow line contribution in
\hb\ is comparable, or
 smaller, than the upper limit for the UV lines (Table 6).
Boroson \& Green (1992) have noted that the maximum possible
narrow component in \hb\ is rarely more than 3\%, which
is consistent with the values found here for \hb.

The small contribution of the narrow lines to the broad line profiles appears
to imply a correspondingly small covering factor of the NLR relative to the
BLR, and a low ionization parameter. We note, however, that this is not
necessarily true if dust is embedded
with the ionized gas in the NLR, as recently shown by Netzer \& Laor (1993).

\bc \section {{\rm PHYSICAL IMPLICATIONS OF THE OBSERVED LINE RATIOS}} \ec

The emission line spectra of quasars in the wavelength range
912~\AA$<\lambda<$1216~\AA\ was first
predicted by Bahcall \& Sargent (1967), based on expected correlations between
the
intensities of lines in the near UV and far UV. Of the 13 ions originally
suggested by Bahcall \& Sargent,
we have detected the following four: O~VI, C~III, N~III, and S~VI.
Bahcall \& Sargent predicted C~III~${\lambda}$977 would be the strongest
emission
line in the 912~\AA$<\lambda<$1216~\AA\ range, followed by \lyb, \lyg\ and
O~VI~${\lambda}$1034. As mentioned above, we find the O~VI flux to be about an
order of magnitude larger than the flux in either C~III, \lyb\ or \lyg.
The other 9 ions, which were not detected here, were predicted by Bahcall \&
Sargent to produce weak, or very weak, lines.

Our
understanding of various physical processes responsible for line emission in
AGNs
 has evolved considerably in the 26 years since the original work of
Bahcall \& Sargent (1967). The most recent photoionization models are described
by Ferland \& Persson (1989), Rees, Netzer \& Ferland (1989), Netzer (1990),
and Ferland {\it et al.} (1992). Our discussion below is based on the results
presented in these papers.
The line ratios predicted by these models depend on the density ($n$),
ionization parameter ($U$, defined in \S 6.3), and column
density of the clouds in the BLR (all these models assume gas with a solar
metallicity and use a slab geometry). We do not attempt to use our measurements
to make a detailed analysis of the physical conditions in the BLR, but rather
point out the
major implications of our emission line measurements.

The \lyb\ line was first predicted to be observable in quasars by Bahcall
(1966).
More recent calculations of the \lyb/\lya\ ratio, based on the ``standard'' BLR
models, are presented  by Kwan (1984). As discussed in \S 5.1.2, we find in our
four quasars \lyb/\lya=0.03$-$0.12.
According to Kwan (1984), the \lyb/\lya\ ratio is determined by the value of
$U\times n$,
i.e. by $L_{\rm ionizing}/R^2$, where $L_{\rm ionizing}$ is the ionizing
luminosity
of the quasar, and $R$ is the distance of the cloud from the ionizing continuum
source. Below we estimate
$L_{\rm ionizing}$ for our objects, and use it to get a direct estimate of the
size of the BLR. $L_{\rm ionizing}$ is estimated assuming the ``standard'' AGN
ionizing continuum shape given in Laor \& Draine (1993, Fig.7), which we scale
by
the observed luminosity $\lambda L_{\lambda}$ at $\lambda=1150$\AA. Assuming
H$_0$=50~km~s$^{-1}$ and $q_0=0$, we get $\lambda L_{\lambda}=(7.9, 2.8, 0.60$
and $4.1)\times 10^{46}$ erg s$^{-1}$, for PKS~0405$-$123, H1821+643,
PG~0953+414 and 3C273. These values imply a total ionizing photon flux of (2.9,
1.1, 0.22 and
1.7)$\times
10^{57}$ photons s$^{-1}$. Using the plot of \lyb/\lya\ as a function of
$n\times U$ in Fig.4 of Kwan (1984) and the ionizing photon flux, we get R=2.3,
1.2, 0.46 and 0.47 pc. These values are somewhat larger
than expected from an extrapolation of the
R=0.1$(L/10^{46}$erg~s$^{-1})^{1/2}$pc relation
suggested at lower luminosities (e.g. Netzer 1990, Peterson 1993). However,
the \lyb/\lya\ calculations of Kwan were carried out for
$U=0.01-0.06$, which is
significantly below the values suggested by O~VI and C~III (see below).
 Kwan also assumed a somewhat different ionizing continuum shape than Laor
\& Draine. It remains to be tested whether the \lyb/\lya\ ratio can be used as
a
useful absolute distance indicator for higher values of $U$.

 The \lyg\ line was deblended from C~III~${\lambda}$977 in PKS~0405$-$123, and
we find \lyg/\lyb$=\sim 0.5$. This ratio is consistent with the
\lyg/\lyb$=\sim 0.35$ prediction of Bahcall (1966) for optically thin gas, with
an additional contribution, having a similar line ratio, due to escape from
larger optical depths (see Kwan 1984).

 The C~III~${\lambda}$977 line was measured for PKS~0405$-$123 and H1821+643.
The C~III/O~VI ratio for these two objects are 0.23 and 0.09. The predicted
C~III/O~VI ratio is close
to one for $U=0.1$, and it decreases to $\sim 0.3$ for $U=1$.
Thus the C~III~${\lambda}$977/O~VI ratio measured here suggests a $U\sim 1$
component in the BLR, if both lines originate from the same cloud distribution.
The C~III~${\lambda}$977/C~III]~${\lambda}$1909 ratio is a good indicator of
the gas density, and is likely to be very weakly dependent on the gas
composition
and column density. We measured C~III/C~III]=0.46 and 0.17 for PKS~0405$-$123
and
H1821+643. These ratios imply $n\sim 3\times 10^9-2\times 10^{10}$~cm$^{-3}$,
if both
lines come from the same distribution of clouds. These values of $n$ are within
the range suggested by the ``standard'' photoionization models.
At higher densities the C~III/C~III] ratio increases since C~III] is
collisionaly
suppressed; for example, at  $n=10^{11}$~cm$^{-3}$ we get C~III/C~III]$>1$,
which is
significantly higher than found here. We note that the C~III~${\lambda}$977
profile is too noisy to determine whether it originates from the same
distribution of clouds as the C~III]~${\lambda}$1909 line.

The N~V~${\lambda}$1240/\lya\ ratio in our sample is more than a factor of two
smaller than found in high-redshift quasars and it displays a very small
scatter,
$0.135\pm 0.01$ (see \S 5.1.3).
The N~V/\lya\ ratio generally increases with $U$ and $n$, however; even for
$n\simeq 10^{10}$~cm$^{-3}$ and $U\simeq 1$ it is still only $\sim 0.05$. As
mentioned in
\S 6.4, the N~V profile is fit very well with a symmetric \lya\ profiles, which
suggests that \lya\ and N~V come from distributions of clouds having similar
$U$ and $n$. (Unless $U$ and $n$ are uncorrelated with velocity).
A possible explanation for the high N~V/\lya\ ratio is a higher than solar
nitrogen abundance and an overall higher than solar metallicity, as recently
suggested by Hamman \& Ferland (1992).
The large systematic errors that are
possible in the measurement of the N~V/\lya\ ratio in low S/N data, and
in particular in high-redshift
quasars (see \S 5.1.3), might affect some of the conclusions of Hamman \&
Ferland (1992) concerning the amount of chemical evolution in quasars.

Earlier studies of high-redshift quasars have noted that the observed
O~VI~${\lambda}$1034/\lya\
ratio requires an optically thin BLR component having a large $U$
 and a covering factor close to unity (e.g. Baldwin \& Netzer 1978). As
mentioned
above (\S 5.1.2) we measured an average O~VI~${\lambda}$1034+\lyb/\lya+N~V
ratio
that is about twice the average ratio found in high-redshift quasars. Our
deblended
average ratio is O~VI~${\lambda}$1034/\lya$=0.32\pm 0.10$ when all the O~VI
flux is
included, and $0.25\pm 0.09$ when we include only the fraction of O~VI which is
fit with the symmetric \lya\ profile. These high O~VI/\lya\ ratios imply a BLR
cloud component with $U\gtsimeq 0.5$ (for $n\sim 10^{10}-10^{11}$~cm$^{-3}$),
which emit the
bulk of the O~VI emission. It is also possible that there is an inner higher
velocity component of the BLR, with a higher $U$, which produces the excess
flux in the wing of O~IV (see \S 6.3). This component has to be optically thin
below the Lyman
limit, so it does not produce a significant \lya\ emission. Ferland,
Korista \& Peterson (1990) and Peterson {\it et al.} (1993) invoked an
optically-thin component to explain non-variable broad wings of \lya\ in a
variable Seyfert 1 galaxy.
 Another possible interpretation for the excess flux in the wing of O~VI is
blendings of other ions, such as Fe~II (\S 6.3).

The He~II~${\lambda}$1640/He~II~${\lambda}$4686 ratio is considered to be a
good
reddening indicator since it is relatively insensitive to the BLR model
parameters. The predicted value for this ratio is in the $8-11$ range, and we
measure values of 21.3, 19.6, 12.4, and 7.3 for PKS~0405$-$123, H1821+643,
PG~0953+414, and 3C273, respectively.
However, the optical and UV spectra were not obtained simultaneously, and
both He~II~${\lambda}$1640 and He~II~${\lambda}$4686 are strongly
blended with other emission components.
 The blending of He~II~${\lambda}$4686 with the optical Fe~II emission was
corrected
using the Boroson \& Green method. This procedure requires a high S/N
emission line template (based on the spectrum of a narrow line quasar),
and it cannot yet be applied for correcting the measurements of
He~II~${\lambda}$1640. The
He~II~${\lambda}$1640/He~II~${\lambda}$4686 ratios measured here could
therefore
include a significant systematic error.

The Si~IV~${\lambda}$1397/O~IV]~${\lambda}$1402 ratio is predicted to be in
the $\sim 1-10$ range. Earlier observations
of high-redshift quasars appeared to indicate Si~IV/O~IV] significantly smaller
than one, but more recent analysis, based on the mean position of the blends
peak indicated Si~IV/O~IV]$\sim 1$. As described above (\S 5.1.5, and Fig.7),
we find using template fitting that Si~IV/O~IV]$\sim 1-3$ is consistent with
our
data for PKS~0405$-$123 and PG~0953+414, and can therefore be explained by
``standard'' BLR models.

The presence of semi-forbidden lines in our spectra,  N~IV]~${\lambda}$1486,
O~III]~${\lambda}$1664, N~III]~${\lambda}$1750, Si~III]~${\lambda}$1892,
and C~III]~${\lambda}$1909, indicates BLR components with densities below
${10}^{10}$ to ${10}^{11}$ cm$^{-3}$. These lines, together with other
prominent
lines, have been pointed out as useful chemical abundance indicators
(e.g. Shields 1976; Uomoto 1984). However, analysis of the chemical abundances
requires detailed modeling of the BLR, which is beyond the scope of this paper.

\bc \section {{\rm SUMMARY}} \ec

We have analyzed the UV emission-line properties of five low-redshift AGNs
using
high-resolution and high-signal-to-noise observations and an objective
algorithmic procedure for modeling and deblending the line profiles. The major
results are the following.

\begin{enumerate}

\item We measure a flux ratio \lyb/\lya\ = 0.03$-$0.12
for the four quasars. Values in this range were first predicted for
quasars using photoionization theory more than a quarter of a century
ago (Bahcall 1966), but previous measurements have been limited by the
 sensitivity and resolution of the available
UV spectra. The \lyb/\lya\ ratios can be used--together with an estimate of
the ionizing luminosity of each object--to calculate directly the size of the
BLR.

\item The cores of the \lya\ and C~IV blends are symmetric in
 two and three objects respectively.
We find flux ratios, Flux(red wing)/Flux(blue wing), that deviate from unity by
less than 2.5\% within 2000~km~s$^{-1}$ of the line peaks. The \lya+N~V
blend  can be fit well using the
``symmetric \lya'' template for both lines. This result
indicates that practically all of the
apparent \lya\ asymmetry at large velocities from the \lya\ line core
is due to blending with N~V. The high degree of
symmetry of \lya\ argues against models in which the velocity field
in the BLR has a significant radial component.

\item The observed
smoothness of the \lya\ and C~IV line profiles requires that at least $\sim
 10^4$ individual
clouds contribute to the broad line emission if bulk velocity is
the only line broadening mechanism.
Electron scattering (e.g. Shields \& McKee 1981; Emmering {\it et al.} 1992)
could explain the smoothness with significantly fewer clouds.

\item The overall similarity of the \lya\ and the C~IV line
profiles rules out models of
the BLR having a distribution of mass-conserving virialized
clouds with $U\propto R^{-1}$.
Models in which  $U$ = constant are consistent
with the similarity of the \lya\ and C~IV profiles.

\item We make the first relatively
accurate measurement of C~III~${\lambda}$977 in quasars. We find
C~III~${\lambda}$977/C~III]~$\lambda1909\simeq$ 0.46 in PKS~0405$-$123
and 0.17
in H1821+642. These ratios imply an electron number density
$n_{\rm e}\sim 3\times 10^9-2\times 10^{10}$~cm$^{-3}$, which is
consistent with ``standard'' photoionization models (e.g. Rees {\it et al.}
1989).

\item An estimate of the \lyg/\lyb\ ratio is possible in PKS~0405$-$123. We
find \lyg/\lyb$\sim 0.5$, which is consistent with the prediction
for optically thin gas (Bahcall 1966), with
an additional contribution, having a similar line ratio, due to escape from
larger optical depths (Kwan 1984).

\item The average deblended O~VI/\lya\ ratio is 0.25$-$0.32. This ratio
and the rather low value found for
the C~III~${\lambda}$977/O~VI line ratio
implies a BLR component with an ionization parameter $U\gtsimeq 0.5$.

\item The maximum contribution of a narrow ([O~III]-like) component to
the observed O~VI, \lya, C~IV,
C~III] and \hb\ line profiles is about $3-6$\% of the total line flux.
The narrow-line contribution is largest in C~III].
The small narrow-line contribution implies either that the covering
factor of the narrow-line region is small or that dust is important (see
Laor \&  Draine 1993).

\item An unresolved component having a FWHM $<230$~km~s$^{-1}$ typically
contributes less than 0.5\% of the observed broad-line flux.

\item The mean N~V/\lya\ ratio for the four quasars is $0.135 \pm 0.01$.
This high value for the N~V/\lya\ ratio may
result from a higher-than-solar N abundance and metallicity (Hamann \& Ferland
1992).

\item Template fitting of the Si~IV/O~IV]$\lambda1400$ blend,
including all the individual multiplet components, indicates Si~IV/O~IV]=$1-3$,
which is close to the values predicted theoretically (e.g. Rees {\it et al.}
1989).
Previous observational estimates in
high-redshift quasars yielded significantly lower values for this
ratio (e.g. Baldwin \& Netzer 1978), which could have
been affected by the intrinsic asymmetry of the Si~IV and the O~IV] multiplets.

\item A comparison of some of the prominent line profiles suggests a dependence
of
$U$ and $n$ on velocity.
In particular, the excess flux in the red wing of O~VI, relative to a fit with
the template \lya\ profile, suggests the
presence of high-velocity optically-thin gas, with $U$ significantly larger
than 1,
which could be located closer to the continuum source than most of the BLR.
The significantly narrower peak of C~III]~${\lambda}$1909, compared with the
peaks
of all other prominent emission lines, suggests a lower density for the
lower-velocity gas, which could be located outside most of the BLR.

\item We measure for the first time in the spectra of individual quasars
the S~VI~$\lambda\lambda$933,945 doublet and the
N~III~${\lambda}$991 emission line.

\item The peaks of the \lya, C~IV, and C~III] emission lines
are typically blueshifted by 150~km~s$
 ^{-1}$ to $250$~km~s$^{-1}$ relative to [O~III]~$\lambda$5007, while
He~II~$\lambda$1640 displays a significantly larger blueshift
of about $500$~km~s$^{-1}$. The low ionization lines of Mg~II, \hb, and
O~I~$\lambda$1304  are in most cases only marginally shifted to the red.
These results are generally similar to what has been found at large redshifts
(cf. Tytler \& Fan 1992).

\end{enumerate}

Emission features are omnipresent in the UV.
We have been unable to find a wavelength range that shows unambiguously
an underlying
featureless continuum. Broad quasi-continuum features at a level of $\sim 10$\%
are evident at most wavelengths.
These emission features can produce systematic errors,
resulting in an overestimate of the true
continuum level and possibly an underestimate of the emission-line
flux.
Another consequence of the high density of UV
emission lines is that essentially
all of the UV lines are blended to some degree.
There is, for example, an unidentified, broad, and in some
cases strong, emission feature at $\lambda\sim 1600$~\AA.
A similar feature has been
detected in the spectra of high-redshift quasars (see e.g.
Boyle 1990).
A practical way to identify the various weak and blended emission
features is by studying narrow-line quasars, where confusion due to blending
is minimized
(e.g. Baldwin {\it et al.} 1988). Spectra of narrow-line quasars can then be
used as templates for the
study of normal quasars, as done by Boroson \& Green (1992) in the optical
regime.

The HST observations presented here provide new information on the
fluxes and profiles of many lines emitted by
different ion species.
These quantitative data (and similar data from other HST observations of
quasars), when combined with detailed photoionization models, will lead
to
improvements in our understanding of the physical conditions, the
chemical abundances, and the dynamics of the gas in the central
regions of active galaxies.

We wish to thank Todd Boroson for obtaining the optical spectrum of H1821+643.
We are grateful for the high quality optical spectra of PKS~0405$-$123 given to
us by B. Wills. We thank R. Blandford, J. Krolik, H. Netzer, D. Maoz, and
B. Wills for helpful advice and suggestions.
This work was supported in part by NASA contract NAG5-1618 and grant GO-2424.01
from the Space Telescope Science Institute, which is operated by the
Association
of Universities for Research in Astronomy, Incorporated, under NASA contract
NAS5-26555. A. L. acknowledges support by NSF grant PHY92-45317.

\newpage
\bc \section*{{\rm APPENDIX}} \ec
\bc \section*{{\rm THE LINE AND CONTINUUM FITTING ALGORITHM}} \ec

Below we describe in details the method we used to model the continuum and the
line
shapes. The advantages of this method are that it is objective, well defined,
and it produces a good fit to the observed profiles.
This method is especially useful for generating an acceptable smooth model for
the
``true'' emission line profiles, as it allows narrow absorption features to be
identified and rejected in the fitting process. The main disadvantage of this
method is that when fitting strongly blended lines (e.g. N~V), the individual
component parameters are not a unique solution, and other solutions
might be more realistic. This disadvantage can be overcome in some cases by
using a
template emission line profile to deblend strongly blended lines, as further
discussed in \S A.4.

\bc \subsection* {A.1. {\em The Continuum Fit}} \ec

To measure lines fluxes, one first needs to define the continuum shape and
level.
It is conventional and convenient for many purposes to describe
the continuum of AGNs as having a power-law shape, i.e.
$d \ln F_{\lambda}/d \ln \lambda=$constant. There is, however, no appreciable
wavelength
range in our spectra which clearly shows such a power-law
continuum uncontaminated with line emission. Weak emission features, with
widths of $\sim 10-100$~\AA,
 are present at a level of
a few percent of the continuum level in many parts of the spectrum. There are
also indications of broader quasi-continuum features of a larger amplitude.
These broader features are suggested by the
fact that a low order polynomial that is fit to the observed continuum at a few
wavelengths generally passes
significantly below (by $\sim 10\%$\ or more) the apparent continuum level at
intermediate wavelengths where no clear emission features are observed (see
also Francis et al. 1991).  These broad features could
be due to low level emission by various blends, in particular Fe II, as
observed in the $\sim 2000-4000$~\AA\ range
(Wills, Netzer \& Wills 1985).  In order to avoid these quasi-continuum
features
as much as possible, we divide each spectrum into sections that are about
$\sim 200$~\AA\ wide (note that rest-wavelengths are used
throughout the Appendix), and for
each section define the local continuum as a power-law that intersects the
spectrum at the two end points.
The flux at each of the two end points is defined as the median flux value
in an interval which is
21 pixels ($\sim 5$ resolution elements) wide,
which corresponds to a wavelength extent $\Delta \lambda \sim \lambda/250$. The
position
of these end-point continuum windows are selected subjectively, under the
guiding principle that they reside in broad local flux minima. The wavelengths
selected for the continuum windows for each object are given in Table 3.

\bc \subsection* {A.2. {\em The Line-Fitting}} \ec

We fit each line blend with a set of Gaussian components.
 We use three Gaussian components, i.e. a total of 9
free parameters, to fit each of the prominent UV lines,
\lya, C~IV~$\lambda$1549, and C~III]~$\lambda$1909, two to three components for
O~VI~$\lambda$1034+\lyb, and a single Gaussian component for all other lines as
they are either too blended with these prominent lines
(e.g. N~V which is blended with \lya),
or too weak (e.g. N~III]~$\lambda$1750) to justify more than one component.
The nonlinear $\chi^2$\ minimization required to make the multiple Gaussian
fit to the data is done using the {\em Levenberg-Marquardt} method, as
implemented by the MRQMIN routine described in {\it Numerical Recipes}
(Press {\it et al.} 1989). The number of Gaussian components used to model
the profile of each line is
the minimum number of components required for an ``acceptable fit", where
an ``acceptable fit" is defined as one which results in a reduced $\chi^2$ of
about 2 or smaller. We found that two Gaussian components are generally not
enough to
provide an ``acceptable fit" for the prominent UV lines, while four components
do not produce a significantly better fit than the one obtained with
three components.

 The major limitation of multiple Gaussian fits is that the best fit solution
is not unique. The specific solution obtained depends on the initial values
for the fit parameters, and it can include physically implausible
parameters, such as a negative flux.
 It is therefore important to have a well defined
procedure to obtain a ``good" initial guess, i.e. initial values which will
lead to
a rapid convergence to an acceptable and stable solution. Below we define the
process by which the initial parameters are obtained.

Note that the observed continuum is contaminated by narrow absorption lines.
These absorption lines have been studied in detail by Bahcall {\it et al.}
(1991, 1992a,b
1993a), and Morris {\it et al.} (1991),
who identified a well defined set of absorption lines. Here, however, we wish
to avoid identifying a region as free of absorption even when it is only
suspected of being absorbed, and we
therefore adopt a lower significance level
for the purpose of identifying spectral regions which are possibly affected by
 absorption.

The line fitting is done according to the following algorithm:
\begin{enumerate}
\item Smooth the data using a Gaussian with $\Sigma=1.75$ pixels (which is
approximately the width of the spectral line spread function). This allows
weak unresolved absorption features to be more easily detected, with a loss of
spectral resolution which is insignificant for our broad
emission line study.

\item Identify regions which are possibly affected by absorption as follows:

\begin{description}
\item a.
Use a median filter with a width of 21 pixels to get an initial
estimate for the local intrinsic emission level $F_{\lambda}^{med}$. This
smoothing effectively eliminates all spectral features narrower than $\sim
600$\ km~s$^{-1}$.
\item b.
Delete all pixels with
 $\Delta<\Delta_{\rm threshold}$, and their neighboring three pixels, where
$\Delta\equiv~(F_{\lambda}^{\rm ob}-~F_{\lambda}^{\rm med})/\sigma$. \
 $F_{\lambda}^{\rm ob} $
is the Gaussian smoothed observed flux, $\sigma$ is the measurement error, and
$\Delta_{\rm threshold}=-2.5$.
\end{description}
\item
Subtract the underlying continuum flux from $F_{\lambda}^{\rm ob}$
to obtain the line flux $F_{\lambda}^l$. The continuum flux is given by
$ F_{\lambda}^c=F_{\lambda_1}^{\rm med}(\lambda/\lambda_1)^{\beta}$,
where $\beta=\ln (F_{\lambda_1}^{\rm med}/F_{\lambda_2}^{\rm med})/
\ln (\lambda_1/\lambda_2)$, and $\lambda_1, \lambda_2$ are the two end points
of
the fitted section of the spectrum.

\item  Obtain initial values for the Gaussian parameters as follows:

\begin{description}
\item a.
Measure the first three moments (i.e. 0th, 1st and 2nd moments) of the flux
distribution within the lower
portion of the line. This portion is defined as
$F_{\lambda}^1\equiv {\rm min}(F_{\lambda}^l, 0.1\times
F_{\lambda}^{\rm max})$, where
$F_{\lambda}^{\rm max}={\rm max}(F_{\lambda}^l)$ for $\lambda_1\le \lambda\le
\lambda_2$. The three moments define the initial values for the first Gaussian
component, which provides an approximate fit to the base of the line.
\item b.
Subtract the first Gaussian component from $F_{\lambda}^l$ to get
$F_{\lambda}^{l'}$.
Measure the first three moments of the flux distribution within the middle part
of the line defined as\\
$F_{\lambda}^2\equiv {\rm min}(F_{\lambda}^{l'}, 0.7\times
F_{\lambda}^{1,\rm max})$,
within $|\lambda-\lambda_0|<20$~\AA,
where
$F_{\lambda}^{1,\rm max}={\rm max}(F_{\lambda}^{l'}$) for
$|\lambda-\lambda_0|<20$~\AA,
and $\lambda_0$ is the wavelength of the observed line peak. These three
moments define the initial values for the second Gaussian components.
\item c.
Subtract the second Gaussian component from the data and measure the first
three
moments of the remaining flux at $|\lambda-\lambda_0|<5$~\AA. These define
the third Gaussian component, which approximately fits the line peak.

\hspace*{-.9cm} This completes the determination of the initial values of the
three Gaussians used to model\\
\hspace*{-.9cm} the profiles of the prominent line.
The initial values for other blended components are\\
\hspace*{-.9cm} determined as follows:
\item d.
Measure the first three moments of $F_{\lambda}^l$ within
$|\lambda-\lambda_i|<10$~\AA\ , where $\lambda_i$\ is the rest wavelength of
the blended line $i$.
\item e.
Repeat d for all additional blended lines (up to three blended lines were
required in
our analysis).

\end{description}

\item Run the nonlinear $\chi^2$\ minimization routine MRQMIN to allow the
initial guess for the Gaussian components
to relax into the best fit solution which minimizes the fit $\chi^2$.
\end{enumerate}

We thus fit each prominent line with three Gaussians which describe the base,
the middle portion, and the core of the line. Note that the value of the fit
$\chi^2$ serves only as a qualitative measure of the
fit and should not be interpreted as indicating the probability of
the fit. This is because our data are characterized by a rather high S/N
($\sim 60$), and
systematic calibration errors of the order of the statistical error
$\sigma$ (mostly determined by the number of photons) are likely to be present
in the spectra. The $\chi^2$ would also be increased if weak absorption
features,
which could not be clearly identified, are common. One should
also note that this fitting procedure cannot fit small amplitude
 narrow features away from the positions of the peaks.

Gaussians do not form a complete set of orthogonal functions and therefore the
specific decomposition obtained here is not unique. The algorithm given above
is, however, well defined, and leads to one specific solution which depends
only on
the given form of $F_{\lambda}^l$.

The nonorthogonality of the individual Gaussian
 components results in a strong covariance of the errors of the fit
parameters. We therefore do not use the $\chi^2$ minimization routine to
quantify the parameter errors, but rather
present the parameters obtained by the algorithm described above as a specific
``acceptable fit". The placement of the continuum level is a possible
source for a large systematic error, and this placement is largely subjective.
To get a rough estimate of the possible
errors introduced by the continuum placement, we repeat the line-fitting
procedure described
above two more times for each blend. In the first iteration we set the
continuum level
at
$F_{\lambda_1}^{\rm med}+\sigma_{\lambda_1}^{\rm med}$ and
$F_{\lambda_2}^{\rm med}+\sigma_{\lambda_2}^{\rm med}$, where
$\sigma_{\lambda_i}^{\rm med}$ is the median of the 21 flux errors in the 21
pixels wide continuum window (note that the formal error in the median is
$\sim \sigma^{\rm med}/\sqrt{21}$, but correlated systematic errors are likely
to be present). In the second iteration,
the continuum is set at $F_{\lambda_1}^{\rm med}-\sigma_{\lambda_1}^{\rm med}$
and $F_{\lambda_2}^{\rm med}-\sigma_{\lambda_2}^{\rm med}$. These two
additional runs are used to estimate the possible range in the values of all
the fit parameters. This range is typically not symmetric around the best fit
parameters, and occasionally the two additional runs give values which deviate
from the best fit value in the same direction.

\bc \subsection*{A.3. {\em Alternative Line-Shape parameterization Methods}}
\ec

The most straightforward parameterization of the line shape is
through the moments of the flux distribution within the line. The
determination of
these moments is subject to a large error beyond the second moment
(the velocity dispersion), as the higher moments are mostly determined by the
far wings of the line where the S/N approaches zero.
Other simple line profile parameterization schemes were described by various
authors,
e.g., Whittle (1985), Boroson \& Green (1992), Emmering {\it et al.} (1992).
However, these various parameterizations are physically meaningful only for
unblended
lines, and such lines are not available here.

A very similar but simpler method to parameterize the line shape is to measure
directly the first three
moments of the flux distribution within the base, middle, and core sections
of the line, as done above but without applying the detailed model fitting
scheme described above (through the $\chi^2$ minimization).
This method yields results which are very similar to the those
obtained above following the model fitting if: 1. the observed profile is
characterized by a high S/N, 2. the observed profile is not significantly
affected by absorption. However, significant narrow absorption features are
present in many
parts of the spectrum, and our line fitting method identifies these regions and
excludes them from the fit. Our method thus allows a more reliable estimate of
the
intrinsic emission line shape parameters.

The non-uniqueness of our Gaussian decomposition method can be overcome if the
line profiles are modeled using a complete set of orthogonal functions.
For example, van der Marel \& Franx (1992) use the first few terms in the
Gauss-Hermite series to quantify
the shape of isolated stellar absorption lines in spectra of galaxies. This
method
is mostly suitable for unblended lines. In our case, all lines are blended,
and an acceptable fit will require a large number of terms in the series
expansion.
Furthermore, the individual expansion terms do not have a straightforward
physical
interpretation when the fit is made to blended lines.

\bc \subsection*{A.4. {\em Line Deblending Using a Template}} \ec

Earlier studies (e.g. Baldwin \& Netzer 1978; Uomoto 1984) have generally
deblended emission lines using the observed profile of one of the prominent
lines,
typically C~IV,  as a
template for the other line profiles. This method has two major drawbacks:
1. It relies on the presence of strong unblended lines, while
as shown below, all strong emission lines are blended to various degrees.
2. The basic premise that all lines have similar profiles is not necessarily
valid.

Some of the emission lines are multiplets of two or more
components (e.g C~IV, N~V, O~VI), and it is not clear whether the apparent
profile
differences reflect intrinsic differences in the velocity distribution of the
line emitting gas. The only way to test the significance of the profile
differences is through
template fitting. The best ad hoc solution we found to the first drawback
mentioned above
is using the blue wing of \lya, and assuming \lya\ is intrinsically exactly
symmetric. The second drawback is avoided since our purpose here is in fact to
test the significance of the profile differences. If a template fit is
determined to be acceptable then it can be used to deblend the parameters (flux
and
velocity shift) of the individual lines.

We fit the various observed emission blend profiles with a sum of ``symmetric
\lya''
templates. We use one template component for each line, or if the line is a
multiplet, one template component for each multiplet component (e.g., two for
C~IV).
The flux normalization and velocity shift of each component are free
parameters.
However, the velocity differences between components of a multiplet is fixed
(e.g. 498.7~km~s$^{-1}$ for C~IV), and the flux ratio in components of a
multiplet
is fixed at either the statistical weight ratio (e.g. 2:1 for C~IV), or at a
ratio
of 1.
We then look for the parameters of the best fit solution, i.e. the solution
which
minimizes the fit $\chi^2$. The $\chi^2$ minimization is done using the MRQMIN,
as
described \S A.2.

When this method is successful in producing an acceptable fit, it most probably
gives a significantly more realistic measurement of the flux in the various
lines
contributing to the emission blend, as compared with the Gaussian components
fittings. We note,
however, that some lines (e.g., C~III]) are often significantly different from
\lya, and in some objects (e.g. H1821+643) all lines are clearly asymmetric. In
all
these cases the template method cannot be applied.


\newpage
\bc \section*{{\rm REFERENCES}} \ec
\begin{description}
\item Atwood, B., Baldwin, J. A., \& Carswell, R. F. 1982, ApJ, 257, 559
\\[-.9cm]
\item Bahcall, J. N. 1966, ApJ, 145, 684 \\[-.9cm]
\item Bahcall, J. N., \& Sargent, W. L. W. 1967, ApJ, 148, L65\\[-.9cm]
\item Bahcall, J. N., Jannuzi, B. T., Schneider, D. P., Hartig, G. F., Bohlin,
R., \& Junkkarinen, V. 1991, ApJ, 377, L5 \\[-.9cm]
\item Bahcall, J. N., Jannuzi, B. T., Schneider, D. P., Hartig, G. F., \&
Green, R. F. 1992a, ApJ, 397, 68 \\[-.9cm]
\item Bahcall, J. N., Jannuzi, B. T., Schneider, D. P., Hartig, G. F., \&
Jenkins, E. B. 1992b, ApJ, 398, 495 \\[-.9cm]
\item Bahcall, J. N., Jannuzi, B. T., Schneider, D. P., \& Hartig, G. F.
 1993a, ApJ, 405, 491 \\[-.9cm]
\item Bahcall, J. N., Bergeron, J., Boksenberg, A., Hartig, G. F., Jannuzi, B.
T.,
Kirhakos, S., Sargent, W. L. W., Savage, B. D., Schneider, D. P., Turnshek, D.
A.,
Weymann, R. J., \& Wolfe, A. M., 1993b, ApJS, in press \\[-.9cm]
\item Baldwin, J. A., McMahon, R., Hazard, C., \& Williams, R. E. 1988, ApJ,
327,
103 \\[-.9cm]
\item Baldwin, J. A., \& Netzer, H. 1978, ApJ, 226, 1\\[-.9cm]
\item Boroson, T. A. \& Green, R. F., 1992, ApJS, 80, 109 \\[-.9cm]
\item Boyle, B. J. 1990, MNRAS, 243, 231 \\[-.9cm]
\item Buson, L. M. \& Ulrich, M. H. 1990, A\&A, 240, 247 \\[-.9cm]
\item Burstein, D. \& Heiles, C., 1982, ApJ, 87, 1165 \\[-.9cm]
\item Capriotti, E., Foltz, C., \& Byard, P. 1981, ApJ, 245, 369\\[-.9cm]
\item Collin-Souffrin, S., Dyson, J. E., McDowell, J. C., \& Perry, J. J. 1988,
MNRAS, 232, 539 \\[-.9cm]
\item Corbin, M. R. 1990, ApJ, 357, 346 \\[-.9cm]
\item Cristiani, S. \& Vio, R. 1990, A\&A, 227, 385 \\[-.9cm]
\item Dumont, A. M., \& Mathez, G. 1981, A\&A, 102, 1 \\[-.9cm]
\item Dumont, A. M., \&  Collin-Souffrin, S. 1990, A\&A, 229, 313 \\[-.9cm]
\item Emmering, R. T., Blandford, R. D., \& Shlosman, I. 1992, ApJ, 385,
460\\[-.9cm]
\item Ferland, G. J., K. T. Korista, \& Peterson, B. M. 1990, ApJ, 363,
L21\\[-.9cm]
\item Ferland, G. J., Netzer, H., \& Shields, G. A. 1979, ApJ, 232,
382\\[-.9cm]
\item Ferland, G. J., \& Persson, S. E. 1989, ApJ, 347, 656 \\[-.9cm]
\item Ferland, G. J., Peterson, B. M., Horne, K., Welsh, W. F. \& Nahar, S. N.
1992, ApJ, 387, 95\\[-.9cm]
\item Francis, P. J., Hewett, P. C., Foltz, C. B., Chaffee, F. H., Weymann, R.
J.,
\& Morris, S. 1991, ApJ, 373, 465  \\[-.9cm]
\item Gaskell, C. M. 1982, ApJ, 263, 79 \\[-.9cm]
\item Gaskell, C. M., Shields, G. A., \& Wampler, E. J. 1981, ApJ, 249, 443
\\[-.9cm]
\item Gondhalekar, P. M., O'Brien, P., \& Wilson, R. 1986, MNRAS, 222, 71
\\[-.9cm]
\item Grandi, S. A., \& Phillips, M. M. 1979, ApJ, 232, 659 \\[-.9cm]
\item Green, R. F., Pier, J. R., Schmidt, M., Estabrook, F. B., Lane, A. L., \&
Wahlquist, H. D. 1980, ApJ, 239, 483 \\[-.9cm]
\item Hamann, F., \& Ferland, G. 1992, ApJ, 391, L53 \\[-.9cm]
\item Kallman, T. R., \& Krolik, J. H. 1986, ApJ, 308, 805 \\[-.9cm]
\item Kallman, T. R., Wilkes, B. J., Krolik, J. H., \& Green, R. 1993, ApJ,
403, 45\\[-.9cm]
\item Kinney, A. L., Bohlin, R. C., Blades, J. C., \& York, D. G. 1991, ApJS,
75, 645 \\[-.9cm]
\item Kinney, A. L., Huggins, P. J., Bergman, J. N. \&  Glassgold, A. E. 1985,
ApJ, 291, 128 \\[-.9cm]
\item Kinney, A. L., Huggins, P. J., Glassgold, A. E. \&  Bergman, J. N. 1987,
ApJ, 314, 145 \\[-.9cm]
\item Kolman, M., Halpern, J. P., Shrader, C. R. \& Filippenko, A. V. 1991,
ApJ,
373, 57 \\[-.9cm]
\item Koratkar, A. P. \& Gaskell C. M. 1991, ApJ, 375, 85 \\[-.9cm]
\item Korista, K. T., Weymann, R. J., Morris, S. L., Kopko, M., Turnshek, D.
A.,
Hartig, G. F., Foltz, C. B., Burbidge, E. M., \& Junkkarinen, V. T. 1993, ApJ,
401,
529 \\[-.9cm]
\item Kriss, G. A., Davidsen, A. F., Blair, W. P., Ferguson, H. C., \& Long, K.
S. 1992, ApJ, 394, L37 \\[-.9cm]
\item Kwan, J. 1984, ApJ, 283, 70 \\[-.9cm]
\item Kwan, J., \& Krolik, J. H. 1981, ApJ, 250, 478 \\[-.9cm]
\item Laor, A., \& Draine, B. T. 1993, ApJ, 402, 441 \\[-.9cm]
\item Maoz D., Netzer, H., Mazeh,T., Beck, S., Almoznino, E., Leibowitz, E.,
Brosch, N., Mendelson, H.,  \& Laor, A. 1991, ApJ, 367, 493\\[-.9cm]
\item Morris, S. L., Weymann, R. J., Savage, B. D., \& Gilliland, R. L. 1991,
ApJ,
377, L21\\[-.9cm]
\item Morton, D. C., 1991, ApJS, 77, 119 \\[-.9cm]
\item Netzer, H. 1990, in  Active Galactic Nuclei (SAAS-FEE Advanced Course 20,
Swiss Society for Astrophysics and Astronomy), ed. Courvoisier T. J. L., \&
Mayor M.
(Berlin: Springer), 57 \\[-.9cm]
\item Netzer, H., \& Laor, A. 1993, ApJ, 404, L51 \\[-.9cm]
\item Netzer, H., Laor, A., \& Gondhalekar, P. M. 1992, MNRAS, 254, 15\\[-.9cm]
\item Netzer, H., \& Wills, B. J. 1983, ApJ, 275, 445 \\[-.9cm]
\item O'Brien, P. T., Gondhalekar, P. M. \& Wilson, R., 1988, MNRAS, 233,
801\\[-.9cm]
\item Osmer, P. A., \& Smith, M. G. 1976, ApJ, 210, 267 \\[-.9cm]
\item Osterbrock, D. E. 1989, Astrophysics of Gaseous Nebulae and Active
Galactic Nuclei, (California, University Science Books)  \\[-.9cm]
\item Peterson, B. M. 1993, PASP, 105, 247 \\[-.9cm]
\item Peterson, B. M., Ali, B., Horne, K., Bertram, R., Lame, N. J., Pogge, R.
W.
\& Wagner, R. M. 1993, ApJ, 402, 469 \\[-.9cm]
\item Press, W. H., Flannery, B. P., Teukolsky, S. A., \& Vetterling, W. T.
1989,
``Numerical Recipes: The Art of Scientific Computing'',
(Cambridge University Press) \\[-.9cm]
\item Rees, M. J., Netzer, H., \& Ferland, G. J., 1989, ApJ, 347, 640 \\[-.9cm]
\item Savage, B. D. {\it et al.} 1993, ApJ, submitted \\[-.9cm]
\item Savage, B. D. \& Mathis, J. S. 1979, ARAA, 17, 73 \\[-.9cm]
\item Schneider, D. P. {\it et al.} 1993, ApJS, in press \\[-.9cm]
\item Seaton, M. J., 1979. MNRAS, 187, 73p \\[-.9cm]
\item Shields, G. A. 1976, ApJ, 204, 330 \\[-.9cm]
\item Shields, G. A. \& McKee, C. F. 1981, ApJ, 246, L57 \\[-.9cm]
\item Sofia, U, J., Bruhweiler, F. C., \& Kafatos, M. 1988, in A Decade of UV
Astronomy with IUE, ESA SP-281, Vol.2, 269 \\[-.9cm]
\item Stark, A. A., Gammie, C. F., Wilson, R. W., Bally, J., Linke, R. A.,
Heiles,
C., \& Hurwitz, M., 1992, ApJS, 79, 77 \\[-.9cm]
\item Steidel, C. C. \& Sargent, W. L. 1987, ApJ, 313, 171 \\[-.9cm]
\item Steidel, C. C. \& Sargent, W. L. 1991, ApJ, 382, 433 \\[-.9cm]
\item Tinggui, W., Clavel, J., \& Wamsteker, W. 1992, in Physics of Active
Galactic Nuclei, eds. W. J. Duschl \& S. J. Wagner (Springer-Verlag, Berlin),
261 \\[-.9cm]
\item Tytler, D., \& Fan, X. M. 1992, ApJS, 79, 1 \\[-.9cm]
\item Ulrich, M. H. 1989, A\&A, 220, 71 \\[-.9cm]
\item Ulrich, M. H. et al. 1980, MNRAS, 192, 561 \\[-.9cm]
\item Uomoto, A. 1984, ApJ, 284, 497 \\[-.9cm]
\item van der Marel, R. P. \& Franx, M. 1993, ApJ, 407, 525 \\[-.9cm]
\item V\'{e}ron-Cetty, M. P., V\'{e}ron, P. 1991, ``A catalogue of Quasars and
Active Nuclei
(5$^{\rm th}$ Edition)'' (Munich: European Southern Observatory) \\[-.9cm]
\item V\'{e}ron-Cetty, M. P., V\'{e}ron, P., \& Tarenghi, M. 1983, A\&A, 119,
69 \\[-.9cm]
\item Whittle, M. 1985, MNRAS, 213, 1 \\[-.9cm]
\item Wilkes, B. J. 1984, MNRAS, 207, 73 \\[-.9cm]
\item Wilkes, B. J. 1986, MNRAS, 218, 331 \\[-.9cm]
\item Wilkes, B. J., \& Carswell, R. F. 1982, MNRAS, 201, 645 \\[-.9cm]
\item Wills, D., \& Netzer, H. 1979, ApJ, 233, 1 \\[-.9cm]
\item Wills, B. J., Netzer, H., \& Wills, D. 1980, ApJ, 242, L1 \\[-.9cm]
\item Wills, B. J., Netzer, H. \& Wills, D. 1985, ApJ, 288, 94 \\[-.9cm]
\item Wills, B. J., Netzer, Brotherton, M. S., Han, M., Wills, D.,
Baldwin, J. A., Ferland, G. J. \& Browne, I. W. A. 1993, ApJ, in press
\\[-.9cm]
\item Young, P., Sargent, W. L. W., \& Boksenberg, A. 1982, ApJS, 48,
455\\[-.9cm]

\end{description}

\newpage
\section*{{\rm FIGURE CAPTIONS}}
{\bf Figure 1}.
The overall spectral shape of the five objects.
Each panel shows the observed flux density {\it vs.} rest wavelength. The data
were corrected for Galactic reddening, and for the purpose of
presentation, the data were smoothed with a median
filter of $\sim 21$ pixels (except near the prominent lines peaks).
This smoothing effectively eliminates all spectral features narrower than
600~km~s$^{-1}$,
in particular the narrow absorption features present shortward
of \lya. The thin line in each panel is the
 composite quasar spectrum of Francis {\it et al.} (1991) vertically shifted to
overlap with the
HST data at 1800~\AA. This composite was obtained from ground-based
observations
of a few hundred quasars at $z\sim 1-2$. Note the overall similarity between
each
spectrum and the composite, and the broad continuum feature at
$\lambda\sim 1600$~\AA\ (especially in the spectrum of H1821+643).  \\[.5cm]
{\bf Figure 2}.
Intercomparison of the five spectra.
All spectra have been smoothed with a Gaussian with the width of the spectral
resolution (see text). For clarity, all points with
a negative
deviation of more than 3$\sigma$, and the \lya\ geocoronal line (its position
marked with Ly$\alpha_{\oplus}$), were eliminated. We mark the positions of
various
lines that are expected to be present based on photoionization models, or that
have been reported in other objects. In some cases, no feature is apparent at
the
marked position. Note the strong Galactic absorption on both sides of
Ly$\alpha_{\oplus}$.
\\[.5cm]
{\bf Figure 3}.
Deblending of the prominent emission lines. The Gaussian smoothed data in each
panel is shown as an
histogram. The nearly horizontal line at the bottom of each panel is the
assumed
continuum level.
The other solid lines show the individual components and the their sum.
The small black triangles indicate points suspected of being affected
by narrow absorption lines; these were not included in evaluating the fit
$\chi^2$.
The value of the fit $\chi^2$ and the number of degrees of freedom are
indicated in each panel (the value of $\chi^2$ should not be interpreted in a
formal
statistical sense, see text). Note the overall smoothness of the cores of
\lya\ and C~IV in most
objects. {\bf a.} PKS~0405$-$123; {\bf b.} H1821+643; {\bf c.} PG~0953+414;
 {\bf d.} 3C273; {\bf e.} Mrk~205.
\\[.5cm]
{\bf Figure 4}.
\lya\ and C~IV blend asymmetries. Each panel displays the ratio of flux density
in the red wing to the flux density in the blue wing of the fitted model to
each blend as a function of velocity
from line peak. H1821+643 and Mrk~205 have very asymmetric line profiles, but
the other objects have rather similar asymmetries. Note that
C~IV is generally more symmetric than \lya\ .
The dashed line indicates the expected ratio from isotropically emitting clouds
distributed spherically symmetrically in both position and velocity. In this
case special relativistic effects enhance the blue wing emission.
\\[.5cm]
{\bf Figure 5}.
Comparison of the model fits to the profiles of the strongest lines.
Each panel displays
the continuum subtracted line blend profiles, normalized to the line peak
flux density,
as a function of velocity from the expected rest frame position of the line
peak.
O~VI is significantly broader than all other lines, \lya\ is
slightly narrower than C~IV in all objects, and C~III] has the narrowest peak
of all
broad lines (except in H1821+643). The low ionization \hb\ line
 is only slightly shifted in all objects, and is generally significantly
asymmetric.
The narrow [O~III]~$\lambda$5007 line of each object is also displayed. The
peak of
this line is used to define zero velocity.
\\[.5cm]
{\bf Figure 6}.
Template fittings to the O~VI, \lya\ and the C~IV blend. All blends were
fitted with a sum of ``symmetric \lya'' profiles (see text). The thin line
histogram is the observed blend profile, the thick line is the best fit model,
and
the dotted line is the fit for the O~VI blend using a 1:1 flux ratio for the
two components of the O~VI doublet, rather than the statistical weight ratio.
The vertical scale was normalized to unity at the peak of the model fit. The
$\chi^2$
of the fit, the number of degrees of freedom, and the velocity range over which
the fit was made are indicated in each panel. In the O~VI fit note the good
match
at $F_{\lambda}/F_{\rm max}\sim 0.5$, and the excess flux in the observed red
wing
of O~VI. In the \lya\ fit note the generally very good match to red wing of
the blend. In the C~IV fit we also included contribution from the
Si~II$\lambda$1531 doublet. Note the poor match near the peak of C~IV, and the
excess
observed flux in the red wing of C~IV.
\\[.5cm]
{\bf Figure 7}.
Template fit to the Si~IV+O~IV]~$\lambda1400$ blend. The Si~IV doublet is
modeled as
a sum of two components, and the O~IV] multiplet as a sum of 5 components. Each
component has
the ``symmetric \lya'' profile, and the ratio of fluxes follows the ratio of
statistical weights (see text). Both lines are assumed to have a zero velocity
shift
with respect to [O III]~$\lambda 5007$. The Si~IV/O~IV] line ratios are
indicated in each
panel.

\end{document}